\documentclass[twocolumn]{aastex62}

\newcommand{\rrat}{r_\mathrm{ratio}}
\newcommand{\Atil}{\tilde{A}}
\newcommand{\Atc}{\tilde{A}_\mathrm{crit}}

\newcommand{\Hm}{H_\mathrm{max}}
\newcommand{\rhm}{\rho_\mathrm{max}}

\usepackage{bm}
\usepackage{xcolor}
\definecolor{green}{RGB}{0,153,0}

\received{July 17, 2018}
\revised{\today}
\accepted{April 5,2019}
\submitjournal{ApJ}

\shorttitle{Maximum Mass Of Differentially Rotating Strange Quark Stars}
\shortauthors{Szkudlarek et al.}

\begin{document}

\title{Maximum Mass Of Differentially Rotating Strange Quark Stars}

\correspondingauthor{Dorota Gondek-Rosi{\'n}ska}
\email{drosinska@astrouw.edu.pl}
\correspondingauthor{Magdalena Szkudlarek}
\email{msz@astro.ia.uz.zgora.pl}

\author{Magdalena Szkudlarek}
\affiliation{Janusz Gil Institute of Astronomy, University of Zielona G{\'o}ra,\\ Licealna 9,65-417, Zielona G{\'o}ra, Poland}

\author{Dorota Gondek-Rosi{\'n}ska}
\affiliation{Astronomical Observatory, University of Warsaw, Aleje Ujazdowskie 4,
00-478 Warsaw, Poland}

\author{Lo\"{\i}c Villain}
\affiliation{Institut Denis Poisson, Universit\'e de Tours - CNRS (UMR 7013), 37200 Tours, France}

\author{Marcus Ansorg}
\affiliation{Theoretisch-Physikalisches Institut,     Friedrich-Schiller-Universit\"at Jena,\\ Max-Wien-Platz 1, D-07743 Jena, Germany}



\begin{abstract}

    We present the first fully relativistic numerical calculations of differentially rotating Strange Quark Stars models for broad ranges of the maximum density and of the degree of differential rotation. Our simulations are performed with the very accurate and stable multi-domain spectral code FlatStar and use the MIT Bag model for describing strange quark matter. Our calculations based on a thorough exploration of the solution space show that the maximum mass of strange stars depends on both the degree of differential rotation and a type of solution, similarly to neutron stars described by a polytropic equations of state. The highest increase of the maximum mass (compared to the value for a non-rotating star) is obtained for models with a low degree of differential rotation. This highest mass is over four times larger than that of the equivalent non rotating configuration. Comparing our results with calculations done for realistic models of neutron stars, we conclude that with the help of differential rotation, strange stars can sustain masses much larger than stars made from nuclear matter for low degree of differential rotation which reinforces the hope of demonstrating, or of ruling out, the existence of strange matter through the observation by gravitational waves, by gamma rays or by neutrinos of the massive material object born from the merger of a compact binary system or during some supernova events. 

\end{abstract}

\keywords{equation of state --- 
          gravitation --- 
          methods: numerical ---
          stars: neutron, quark, strange --- 
          stars: rotation}


\section{Introduction} 
\label{introduction}

The first detection of gravitational waves (GW) from coalescing neutron stars in a binary system, GW170817, by the Advanced Virgo and LIGO collaboration \citep{A2017} proved that we have a new tool to investigate the properties of matter at extreme densities and, as a consequence, to put constraints on the equation of state (EOS) of compact stars, since both the electromagnetic and the GW signals are strongly affected by the specific nature of their source.
The merger of two neutron stars could lead to the formation of a hot, massive differentially rotating remnant-a neutron star or even a strange quark star \citep{D2016,2019PhRvL.122f1101M} or to a black hole. The outcome mainly depends on the EOS, the total mass and the mass ratio of compact stars in a binary system \citep[see][for a review]{BR2017} and has important implications on
the gravitational wave signal of such a merger. Many attempts have been made to interpret the observations of GW170817 \citep{Rezzolla2018ApJ, Ruiz2018PRD, Shibata2017PRD, 2017SF, 2019PhRvL.122f1101M} but there is still no consensus what is the fate of the post-merger object.

In our paper we investigate properties of differentially rotating strange quark stars, in particular we focus on their maximum masses. 

The possibility of the existence of stable deconfined quark matter was indeed predicted in the early seventies by \citet{bodmer}. Later, the hypothesis that \emph{strange (quark) matter}, with a non-vanishing strangeness per unit baryon number, could constitute the absolute ground state of matter at zero pressure and temperature was made famous by \citet{witten}. He pointed out that three-quark-flavored baryonic matter could be energetically favorable compared to the two-flavored one, meaning that at zero pressure strange quark matter could have a lower energy per baryon than iron $^{56}\mathrm{Fe}$. If that assumption was correct, normal nuclei could decrease their energy by transforming into strange matter \citep[e.g.][]{BD2000}. Among the potential environments that could result in the actual formation of strange quark matter, the early Universe and the interior of compact stars are the most likely, which naturally led to the theoretical study of strange nuclei called \emph{strangelets} \citep{FJ} but also of \emph{strange quark stars} (SQS) entirely built of strange quark matter with a small addition of electrons \citep{AFO, HZS}.

Due to its composition quite different at the microscopic level from that of a usual neutron star (NS), a SQS is macroscopically described by a rather distinct EOS, which implies easily distinguishable properties. For instance, while NSs have a strong density contrast between their center and their surface, the density profile of SQSs is almost constant with a non-vanishing surface value. This feature means that even if static SQSs and NSs can sustain a quite similar maximum mass \citep[whose specific value depends on the precise EOS, see][for a review]{CHZF}, it is no longer the case when the stars are rotating. Indeed, rotation can stabilize stars with masses larger than the maximum mass of non-rotating star~$M_{\mathrm{max,stat}}$ \citep{BSS} and even leads to the occurrence of configurations that cannot exist in the static case. However, while it was shown that for rigidly rotating NSs the maximum allowed mass can be 14\% - 22\% larger than $M_{\mathrm{max,stat}}$ \citep{CST92,CST94a,CST94b}, it is up to 44\% larger for SQSs \citep{GHLPBM, GBZGRDD}.

Such a difference can be crucial in deciding whether SQSs exist, and consequently in learning more about nuclear matter at high density, as the value of the maximum mass of a relativistic star helps to determine if an observed compact object is a black hole or a material celestial body. Moreover, this quantity is also a key-factor for establishing the life span of the short-lived remnant born from the merger of a compact binary system or of a hot neutron star born from a supernova explosion, situations in which the estimation of the maximum mass is further complicated by the fact that such objects are differentially rotating \citep{BSS,SU,STU,GRS11}.

The maximum mass of differentially rotating neutron stars has already been the topic of many studies \citep[e.g.][]{BSS,LBS,MBS,BSB2018,WMR2018,U2017}, but was more recently approached with the relativistic, highly accurate double-domain pseudo-spectral FlatStar code \citep[based on the so-called ``AKM-method'',][]{AKM} which led to several noticeable developments. Indeed, the AKM-method, which was formerly shown to enable the calculation of very extremal configurations of rigidly rotating relativistic stars \citep{AKMb,SA} or rings \citep{A,AP}, allowed a deeper understanding of the general structure of the solution space for constant density stars and $N=1$ (equivalently $\Gamma=2$) polytropes \citep{AGV}, demonstrating for instance the existence of new types of configurations that were not considered in previous studies. One of the key features of the code that made such a thorough exploration possible is the resort to the Newton-Raphson method with the help of which any physical quantity can be used either as a free parameter to vary or as a fixed one.

This property turned out to be crucial for the works \citet{GKVAK} and \citet{SKGVA} where for the first time, the maximum mass and various other astrophysical parameters were calculated for all types of differentially rotating neutron stars modelized by polytropes of various adiabatic indices. It was found that the maximum mass of differentially rotating neutron stars depends on the stiffness of the equation of state, on the degree of differential rotation but also on the type of configuration. More precisely, the highest increase with respect to the maximum mass for non rotating stars with the same equation of state was reached for a modest degree of differential rotation and a moderate stiffness.

In this article, we present the first fully relativistic calculation of the structure of stationary and axisymmetric models of  differentially rotating strange quark stars for broad ranges of the maximum density and of the degree of differential rotation, focusing on the determination of their maximum mass. The paper is organized as follows. First, we briefly describe the numerical model (chosen EOS, rotation law, etc.), sending the reader to previous articles \citep{AGV,GKVAK} for more details on the algorithms. Then, in Section~\ref{sec:test}, we summarize some numerical tests (consistency checks and comparisons with other codes) that were done to validate our results. The latter are displayed in Section~\ref{sec:res}, emphasis being put on the comparison with the values obtained for realistic EOSs of NSs by \citet{MBS}. Finally, Section~\ref{sec:conc} provides a summary of our work and a discussion of its possible extensions.

\section{Numerical model and assumptions}

The FlatStar code, based on spectral methods and on the Newton-Raphson algorithm \citep{AGV,GKVAK,SKGVA}, relies on the assumptions that spacetime and the distribution of matter are axisymmetric and stationary. Additionally, the star is supposed to be made from a perfect fluid which, as far as thermodynamics is concerned, is fully characterized by an EOS giving the pressure in the fluid frame $p$ as a function of the total energy density in that same frame~$\epsilon$ (or equivalently the total mass density $\rho=\epsilon/c^2$, with $c$ the speed of light).

In the case of strange matter, it was shown \citep{GBZGRDD, Z} that, even if the physics of quarks at high density is highly non-linear and complicated, the EOS could be very precisely approximated by a linear relation
\begin{equation}
p = ac^2 \left(\rho - \rho_0 \right)\,, \label{Eq_of_state}
\end{equation}
where $a$ and $\rho_0$ are constants such that $c\sqrt{a}$ is the speed of sound and $\rho_0$ is the density at zero pressure (and as a consequence also at the surface of the SQS in the absence of a crust, as assumed in the following). In this work, as a first step in the study of differentially rotating SQSs, we only consider the simplest form of the MIT Bag model, in which $a = 1/3$ and $\rho_0 = 4.2785 \times 10^{14} \mathrm{g/cm^{3}}$.

Once given a barotropic EOS $p=p(\rho)$, a configuration of a rigidly rotating relativistic star is uniquely determined by solving the axisymmetric and stationary Einstein equations taking into account appropriate boundary conditions but also providing a central (or maximal) mass (or energy) density $\rho$ and an angular velocity $\Omega$. In the case of differential rotation, the situation is more intricate and one can freely choose a rotation law relating a kind of specific relativistic angular momentum to the local angular velocity. As in \citet{AGV,SKGVA,GKVAK}, we adopt here the simple but realistic and astronomically motivated law introduced by \citet{KEH} and considered by many authors \citep[among which][]{CST92,BSS,LBS}. With that law, a rotation profile is specified by $\Omega_c$, the limit of the angular velocity close to the rotation axis, and a parameter which describes the degree of differential rotation. In the present article as in previous ones, we follow \citet{BSS} and use for the latter the dimensionless parameter $\Atil$, whose exact definition can be found in \citet{AGV}. Notice that
\begin{itemize}
\item the \citet{KEH} law implies an angular velocity that monotonously decreases from the axis of rotation to the equator; 
\item $\Atil=0$ corresponds to rigid rotation, $\Omega=\Omega_c$ in the whole star;
\item for $\Atil=1$ the angular velocity at the equatorial surface is around half $\Omega_c$;
\item the degree of differential rotation is an increasing and monotonic function of $\Atil$.
\end{itemize}

Another key difference between rigidly and differentially stationary rotating stars is that the solution space for the latters is much richer. Hence, to uniquely specify a configuration, it is not always sufficient to set the value of the (e.g.) central density, and some geometrical quantities turn out to be quite convenient. In the following, we shall refer to the ratio between the polar and equatorial radii $\rrat=r_p/r_e$, with $0<\rrat\le1$ as we consider only stars without a hole at their center, and also to the so-called "rescaled shedding parameter" $\tilde{\beta}$, see the precise definition in \citet{AGV}, which is such that $0<\tilde{\beta}<1$, with $\tilde{\beta} \rightarrow1$ when a star enters into the toroidal regime ($\rrat=0$), $\tilde{\beta} \rightarrow 0$ at the mass-shedding limit and $\tilde{\beta}=1/2$ for a non-rotating spherical star ($\rrat=1$).

The last technical point worth mentioning is that the FlatStar code uses as a primary thermodynamical variable not the mass density $\rho$, but the corresponding dimensionless relativistic enthalpy~$H(\rho)$ defined by the differential relation
\begin{equation}
dH = \frac{dp}{c^2\,\rho(p)\,+\,p}\,,
\end{equation}
with the property $H(p=0)=0$. As a consequence, in the following our results are equivalently described for given values of the maximum mass density in the star, $\rhm$, or of the maximum enthalpy $\Hm$. Notice that with the EOS~(\ref{Eq_of_state}), one has the relation
\begin{equation}
\rho(H) = \frac{\rho_0}{1+a}\,\left(e^{H\,(a+1)/a}\,+\,a\right)\,,
\end{equation}
so that $H=0$ corresponds to $\rho=\rho_0$, another consequence of which being that in the following sections, a linear scale for $H$ on one axis of a figure is equivalent to a logarithmic-like scale for $\rho$ on the same axis.

\section{Numerical tests}
\label{sec:test}

\subsection{FlatStar testing}

\begin{figure}
\centering
  \includegraphics[width=0.7\linewidth, angle=-90]{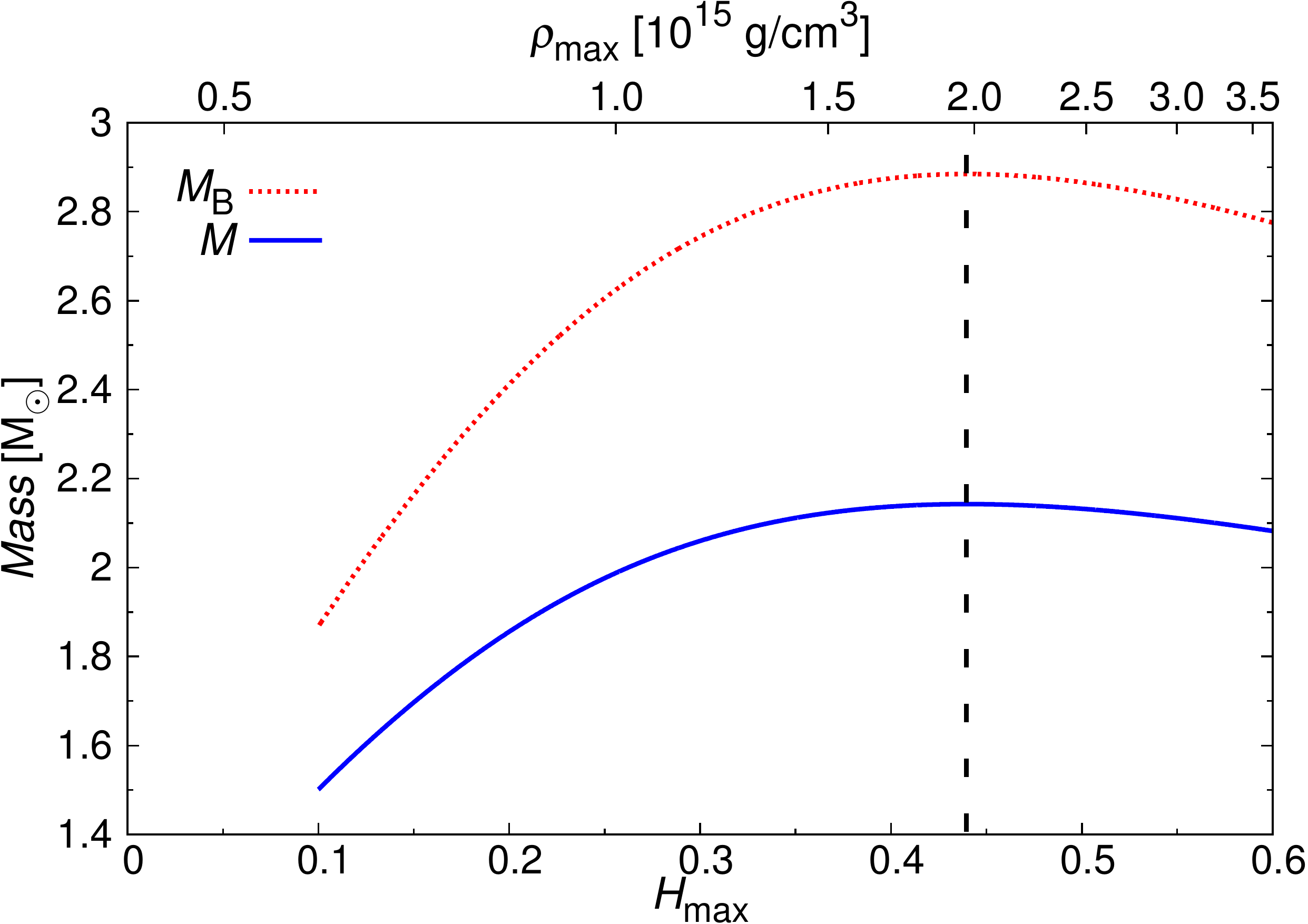}
  \caption{Gravitational (blue solid line) and baryonic (red dotted line) masses (respectively $M$ and $M_B$) as functions of the maximum enthalpy $\Hm$ (bottom axis) and maximum density $\rhm$ (top axis) for sequences of rigidly rotating Strange Quark Stars with fixed angular momentum $J = 2.16 \ [G\mathrm{M_\odot^2}/c]$. The black dashed line identifies the maximum mass configuration. Both maxima are reached at $\Hm = 0.43971$.}
  \label{Fig:J-00015}
\end{figure}

The first internal consistency test of the code we did simply consisted in verifying, for rigidly rotating stars, the agreement of calculated configurations with the first law of thermodynamics for relativistic stars \citep[see e.g.][]{FS}, which implies that in a sequence with fixed angular momentum~$J$ the maxima of gravitational $M$ and baryonic $M_B$ masses occur for the same value of the maximum density $\rhm$ (equivalently of the maximum enthalpy $\Hm$). On Figure~\ref{Fig:J-00015}, we display the results for that test in the case $J = 2.16 \ [G\mathrm{M_\odot^2}/c]$. It can be seen that indeed, with a high degree of precision, the maxima of $M_B$ and of $M$ are reached for the same maximum enthalpy $\Hm = 0.43971$.

\begin{figure}
\centering
    \includegraphics[width=0.7\linewidth, angle=-90]{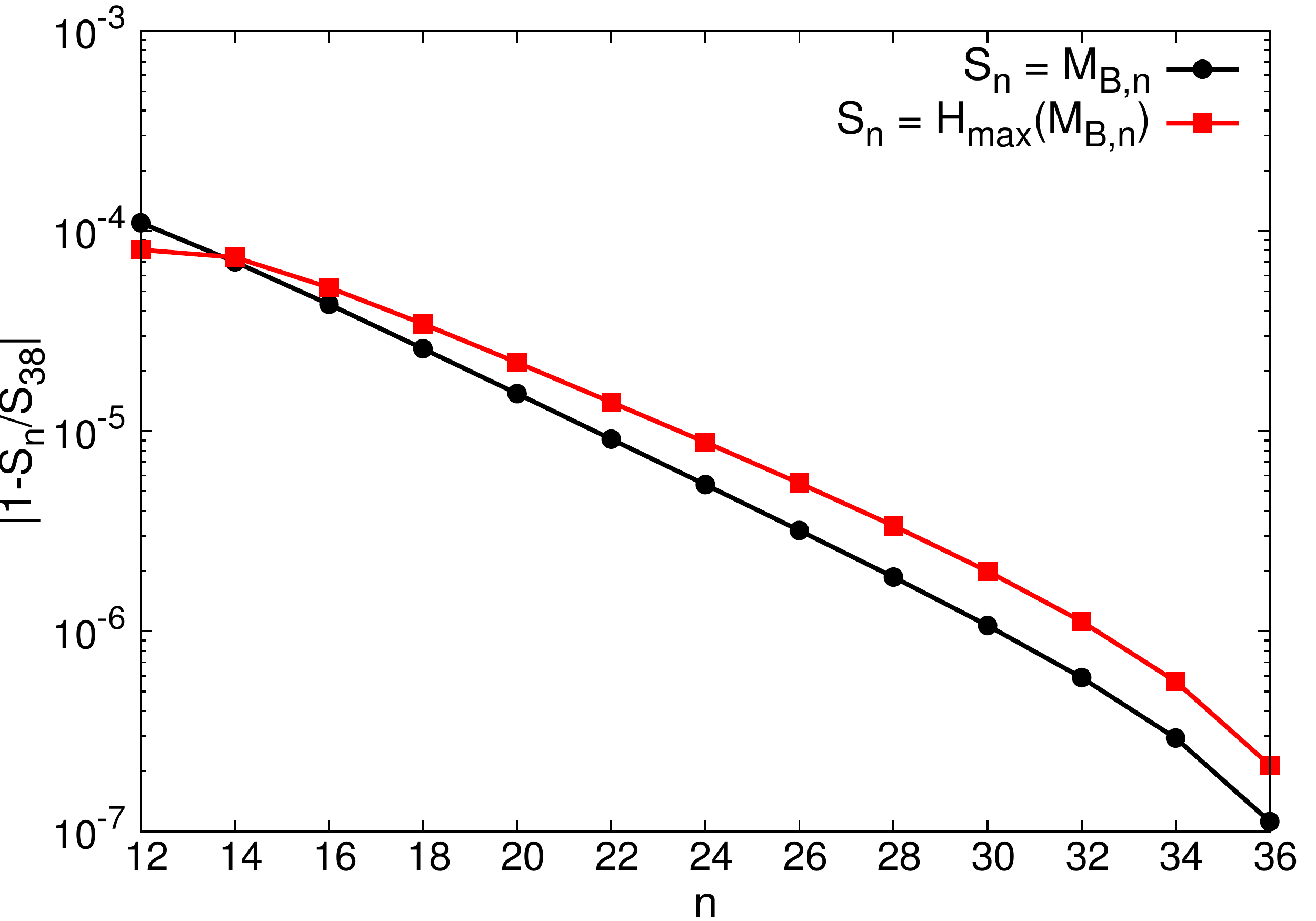}
    \caption{Geometrical convergence rate for the maximum baryonic mass $M_{B, max}$ (black) and the corresponding maximum enthalpy $\Hm$ (red) for differentially rotating Strange Quark Stars with a fixed degree of differential rotation $\tilde{A} = 0.1$. The plot displays the relative accuracy of the $n$th spectral approximation with respect to the accuracy of the $38$th one.}
    \label{Fig:M0max_vs_GridPoints}
\end{figure}

The second test we performed concerned the dependence of the accuracy of the code on the number of grid points [see also the Appendix in \cite{GKVAK}] for differentially rotating stars. One way to proceed was to search for the maximum mass of stars with a fixed degree of differential rotation ($\Atil$) and to check how its value changes with an increasing number of grid points. The results for SQSs with $\Atil = 0.1$ are displayed on Figure~\ref{Fig:M0max_vs_GridPoints}. The relative accuracy of the maximum baryonic mass $M_\mathrm{B}$ (and of the corresponding $\Hm$) obtained for 28 grid points and for 38 points is $10^{-6}$, which is sufficient for our study taking into account all the simplifications made in the model.
As a high number of grid points results in a longer computing time (up to several hours for one single configuration) without a noticeable improvement of the precision, most of the results presented in the following rely on calculations made for 28 grid points.

\subsection{FlatStar, LORENE, RNS - comparison}

\subsubsection{Static and rigidly rotating strange stars}

Finally, we compared the FlatStar code with other publicly available codes allowing the numerical calculation of the structure of static and rigidly rotating stars and with which were made the first fully relativistic calculations for rigidly rotating strange quark stars described by the MIT Bag model \citep{GHLPBM, SKB99}. The first one, RotSeq, is part of the numerical library LORENE [Langage Objet pour la RElativit\'e Num\'eriquE, \href{http://www.lorene.obspm.fr}{http://www.lorene.obspm.fr}, \citet{GLorene}], developed by a numerical relativity team from Paris Observatory in Meudon and based on spectral methods \citep{BA}. The second code, RNS, was written by Nikolaos Stergioulas and constructs models of rigidly rotating relativistic compact stars following the Komatsu {\it et al.}~(1989) method, in which the field equations are converted into integral equations using appropriate Green’s functions \citep{SF95}.

We started with static (non-rotating) strange star configurations, searching for the one with maximum mass. All three codes were found to be in very good agreement, the maxima being associated to the same value of the central density~$\rho_{\mathrm{c}} = 2.059 \cdot \mathrm{10^{15} g \ cm^{-3}}$ (central enthalpy $H_{\mathrm{c}} = 0.4514$). To be more specific, the discrepancies between the values of the calculated maximum gravitational mass~$M_{\mathrm{max}}^{\mathrm{stat}}$ were only of 0.005\% and 0.2\% with LORENE and with RNS, respectively (see Table~\ref{Tab:Comparison}).

In the case of rigid rotation, getting a precise value of the maximum mass was more difficult, mainly due to the lower accuracy of the RNS code. In order to get a meaningful comparison of all codes, we consequently adopted the strategy already used in \citet{SKB99} (Table~2 in the article): we looked for the configuration with maximum mass using LORENE, and then performed calculations for stars with the corresponding $\rho_c$ using the RNS and FlatStar codes. The differences between the outputs of the codes were larger than for static stars but still in good agreement: depending on the physical quantity, the values differ of about 0.1\% - 1.0\% with LORENE and of 0.02\% - 1.1\% with RNS, as can be seen in the second part of Table~\ref{Tab:Comparison}.

\begin{deluxetable*}{lllllllllll}
\tablecaption{Comparison of three codes (FlatStar, RotSeq and RNS) for non-rotating and rigidly rotating Strange Quark Stars [RotSeq - \citet{GHLPBM}; RNS - \citet{SKB99}]. For all configurations, the Table displays the gravitational mass $M$, the baryonic mass $M_\mathrm{B}$, the angular velocity $\Omega$, the central mass density $\rho_\mathrm{c}$, the central enthalpy $H_\mathrm{c}$, the circumferential radius $R_\mathrm{circ}$, the coordinate equatorial radius $r_e$, the ratio between the polar and equatorial radii $\rrat=r_p/r_e$, the angular momentum $J$ and the ratio between the kinetic and potential energies $T/|W|$. The upper part of the Table contains static configurations ($\Omega=0$) with maximum mass, while the lower part presents rigidly rotating configurations with a central mass density $\rho_{rm c}$ such that the star is the one with maximum mass for the RotSeq code. As explained in the text, those stars are not exactly those of maximum mass for RNS and FlatStar. \label{Tab:Comparison}}
    \tablehead{
      \colhead{}&
      \colhead{$M [\mathrm{M_\odot}]$}&
      \colhead{$M_{\mathrm{B}} [\mathrm{M_\odot}]$}&
      \colhead{$\Omega [\mathrm{rad \ s^{-1}}]$}&
      \colhead{$\rho_{\mathrm{c}} [\mathrm{10^{15} g \ cm^{-3}}]$}&
      \colhead{$H_{\mathrm{c}}$}&
      \colhead{$R_{\mathrm{circ}} [\mathrm{km}]$}&
      \colhead{$r_e [\mathrm{km}]$}&
      \colhead{$\rrat$}&
      \colhead{$J [G\mathrm{M_\odot^2}/c]$}&
      \colhead{$T/|W|$}\\
      \hline
      \multicolumn{11}{c}{$M_{\mathrm{max}}^{\mathrm{stat}}$}
      }
      \startdata
      FlatStar & 1.9637 & 2.6489 & 0 & 2.059 & 0.4514 & 10.71 & 7.532 & 1 & 0 & 0 \\
      RotSeq   & 1.9638 & 2.6255 & 0 & 2.059 & 0.4514 & 10.71 & 7.532 & 1 & 0 & 0 \\
      RNS      & 1.9676 & 2.6330  & 0 & 2.059 & 0.4514 & 10.71 & 7.527 & 1 & 0 & 0 \\
      \hline
      \multicolumn{11}{c}{$M_{\mathrm{max}}^{\mathrm{rig}}$} \\
      \hline
      FlatStar & 2.8188 & 3.778  & 9560.19 & 1.261 & 0.32 & 16.463 & 11.308 & 0.4643 & 7.013 & 0.209 \\
      RotSeq   & 2.8310  & 3.751  & 9547.0  & 1.261 & 0.32 & 16.54  & 11.37  & 0.4618 & 7.084 & 0.21  \\
      RNS      & 2.8339 & 3.759  & 9562.61 & 1.261 & 0.32 & 16.425 & 11.263 & 0.466  & 7.093 & 0.21  \\
     \enddata
\end{deluxetable*}

\subsubsection{Differentially rotating stars}

\begin{figure}
\centering
  \includegraphics[width=0.7\linewidth, angle=-90]{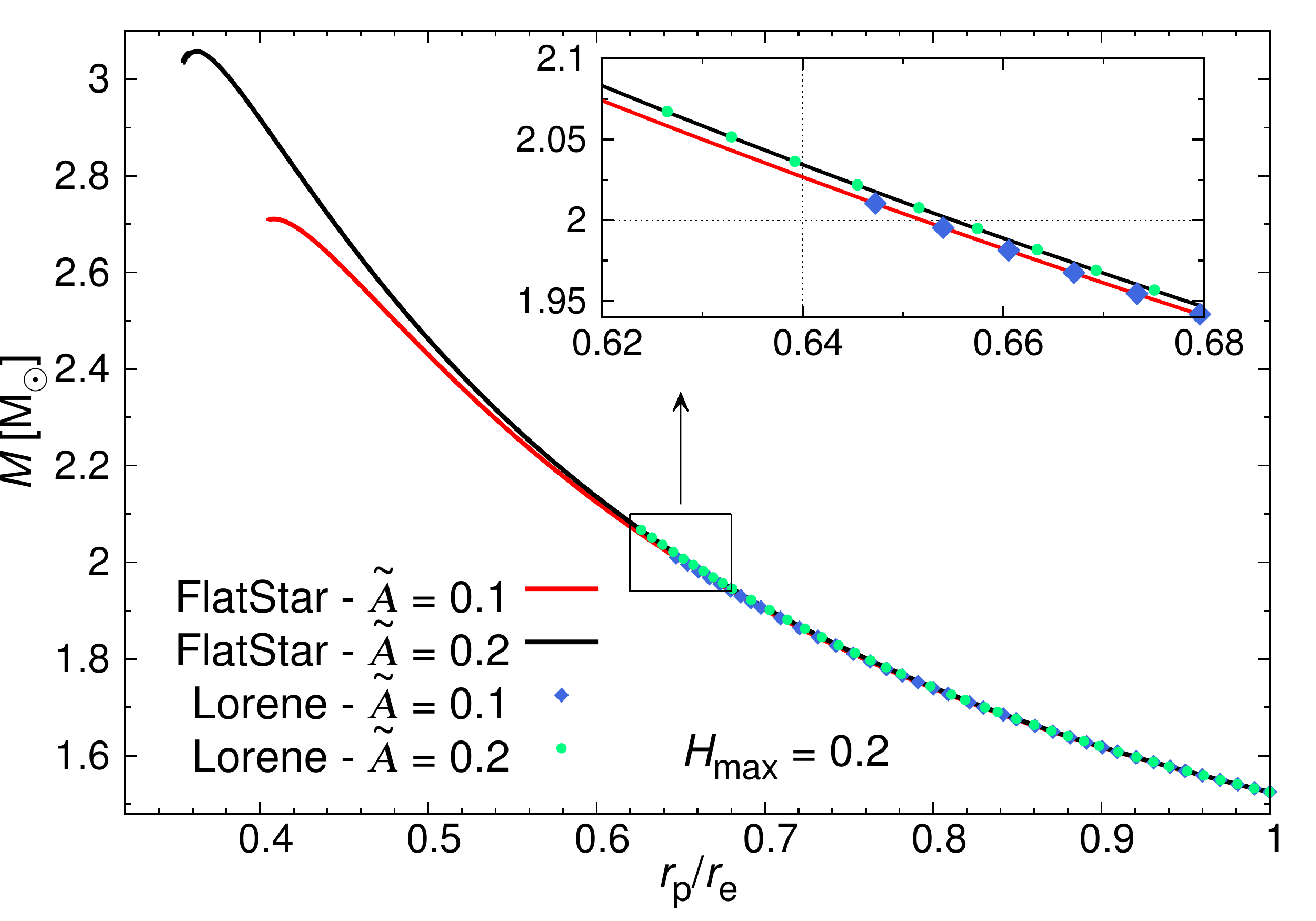}
  \caption{Gravitational mass $M$ as a function of the ratio between polar and equatorial radii $\rrat=r_p/r_e$ for sequences of differentially rotating Strange Quark Stars calculated with FlatStar (solid lines) and LORENE (dots or diamonds), for a fixed maximum density (maximum enthalpy) $\rho_{\mathrm{max}} = 0.82 \cdot 10^{15} \mathrm{g/cm^3}$ ($\Hm = 0.2$) and two values of the degree of differential rotation $\tilde{A} = 0.1$ and $0.2$. As can be seen, while the agreement between the two codes is very good for slow rotation, LORENE fails to converge for rapidly rotating configurations characterized by the lowest values of $\rrat$.}
  \label{Fig:FlatStar_vs_Lorene_Hmax-02}
\end{figure}

\begin{figure}
\centering
  \includegraphics[width=0.7\linewidth, angle=-90]{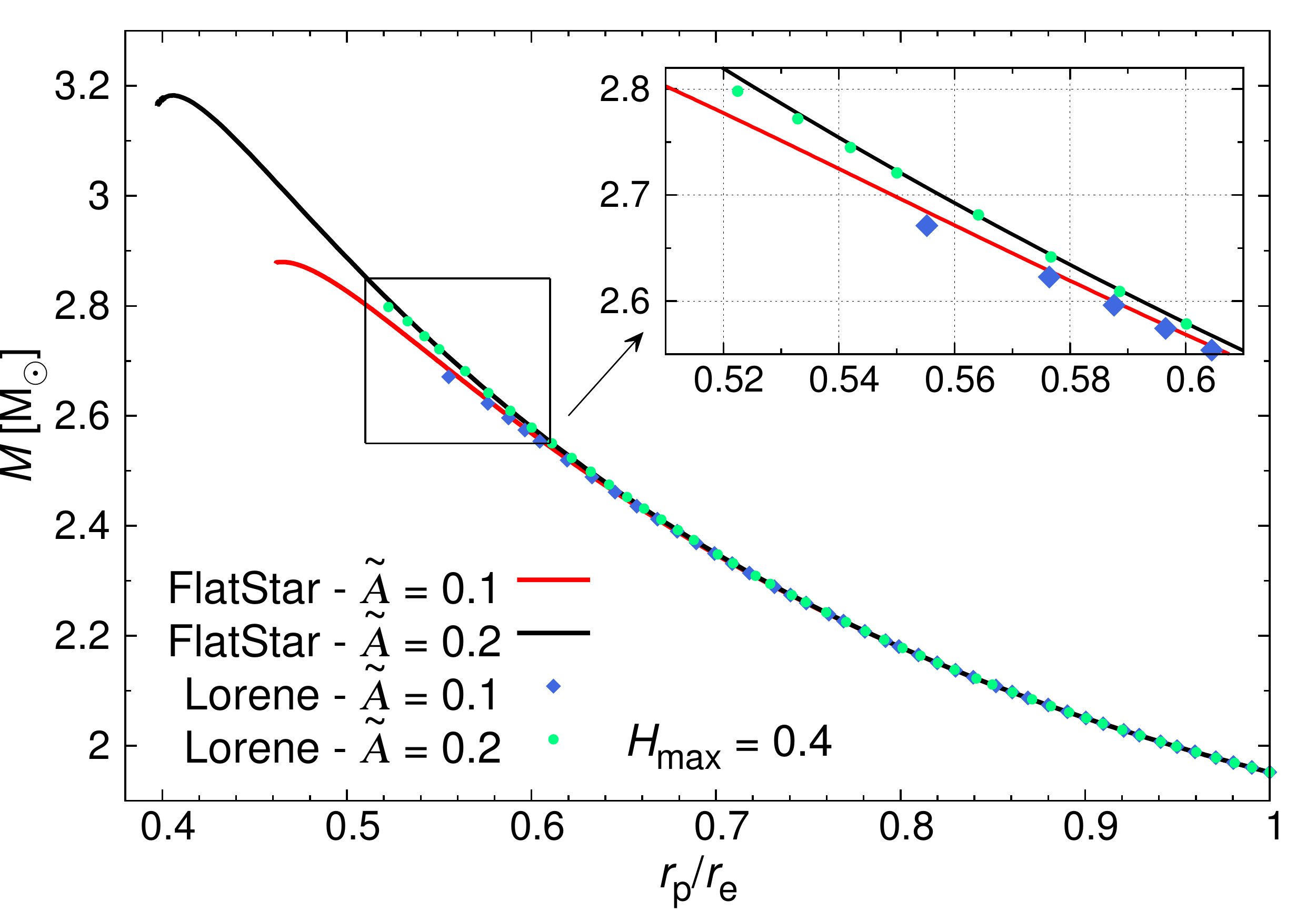}
  \caption{Same as Fig.~\ref{Fig:FlatStar_vs_Lorene_Hmax-02} for stars with maximum density (maximum enthalpy) $\rho_{\mathrm{max}} = 1.70 \cdot 10^{15} \mathrm{g/cm^3}$ ($H_{\mathrm{max}} = 0.4$).}
  \label{Fig:FlatStar_vs_Lorene_Hmax-04}
\end{figure}

\begin{table*}
\centering
    \caption{Comparison of differentially rotating Strange Quark Stars obtained with FlatStar and with LORENE, for two values of $\rrat$ ($0.9$ and $0.7$) as well as of $\Atil$ ($0.1$ and $0.2$), and for fixed $\rhm = 0.82 \cdot 10^{15} \mathrm{g/cm^3}$ ($\Hm = 0.2$). In the Table are displayed the gravitational mass $M$, the baryonic mass $M_B$, the circumferential radius $R_{circ}$, the angular momentum $J$, the central angular velocity $\Omega_c$, the corresponding central rotational frequency $f_c=\Omega_c/(2\pi)$ and the ratio between the kinetic and potential energies $T/|W|$.}
    \begin{tabular}{@{}clllll@{}}
    \hline
    \hline
    &  & \multicolumn{2}{c}{ $\tilde{A} = 0.1$} & \multicolumn{2}{c}{ $\tilde{A} = 0.2$} \\
    &  & FlatStar & LORENE & FlatStar & LORENE \\
    \hline
					   & $M [\mathrm{M_\odot}]$		         & 1.61758 & 1.61767 & 1.61839 & 1.61853 \\
					   & $M_{\mathrm{B}} [\mathrm{M_\odot}]$ & 2.09699 & 2.07821 & 2.09828 & 2.07954 \\
                       & $R_{\mathrm{circ}} [\mathrm{km}]$	 & 11.4927 & 11.4938 & 11.4857 & 11.4871 \\
    $\rrat=0.9$		   & $J [G\mathrm{M_\odot^2}/c]$	     & 0.93263 & 0.93272 & 0.93015 & 0.93051 \\
					   & $\Omega_c [\mathrm{rad \ s^{-1}}]$  & 4178.87 & 4178.44 & 4327.73 & 4328.22 \\
					   & $f_c [Hz]$				             & 665.087 & 665.019 & 688.779 & 688.858 \\
					   & $T/|W|$				             & 0.03338 & 0.03338 & 0.03317 & 0.03318 \\
    \hline
					   & $M [\mathrm{M_\odot}]$		         & 1.90198 & 1.90213 & 1.90668 & 1.90681 \\
					   & $M_{\mathrm{B}} [\mathrm{M_\odot}]$ & 2.47319 & 2.45111 & 2.48060 & 2.45835 \\
                       & $R_{\mathrm{circ}} [\mathrm{km}]$	 & 12.9953 & 12.9969 & 12.9561 & 12.9585 \\
    $\rrat=0.7$		   & $J [G\mathrm{M_\odot^2}/c]$	     & 2.38766 & 2.38804 & 2.38863 & 2.38921 \\
					   & $\Omega_c [\mathrm{rad \ s^{-1}}]$  & 7111.69 & 7110.62 & 7421.57 & 7421.32 \\
					   & $f_c [Hz]$				             & 1131.86 & 1131.69 & 1181.18 & 1181.14 \\
					   & $T/|W|$				             & 0.11146 & 0.11145 & 0.11082 & 0.11086 \\
    \hline
    \end{tabular}
\label{Tab:FlatStar_vs_Lorene_Hmax-02}
\end{table*}

\begin{table*}
\centering
    \caption{Same as in Tab.~\ref{Tab:FlatStar_vs_Lorene_Hmax-02} but for $\rhm = 1.70 \cdot 10^{15} \mathrm{g/cm^3}$ ($\Hm = 0.4$).}
    \begin{tabular}{@{}clllll@{}}
    \hline
    \hline
    &  & \multicolumn{2}{c}{ $\tilde{A} = 0.1$} & \multicolumn{2}{c}{ $\tilde{A} = 0.2$} \\
    &  & FlatStar & LORENE & FlatStar & LORENE \\
    \hline
					   & $M [\mathrm{M_\odot}]$		         & 2.05075 & 2.05085 & 2.05070 & 2.05081 \\
					   & $M_{\mathrm{B}} [\mathrm{M_\odot}]$ & 2.76025 & 2.73548 & 2.76040 & 2.73566 \\
                       & $R_{\mathrm{circ}} [\mathrm{km}]$	 & 11.3862 & 11.3873 & 11.3710 & 11.3722 \\
    $\rrat=0.9$		   & $J [G\mathrm{M_\odot^2}/c]$	     & 1.49360 & 1.49381 & 1.48696 & 1.48721 \\
					   & $\Omega_c [\mathrm{rad \ s^{-1}}]$  & 5250.26 & 5250.07 & 5510.20 & 5510.08 \\
					   & $f_c [Hz]$				             & 835.604 & 835.575 & 876.976 & 876.956 \\
					   & $T/|W|$				             & 0.03501 & 0.03501 & 0.03476 & 0.03477 \\
    \hline
					   & $M [\mathrm{M_\odot}]$		         & 2.34800 & 2.34792 & 2.35005 & 2.34975 \\
					   & $M_{\mathrm{B}} [\mathrm{M_\odot}]$ & 3.15643 & 3.12785 & 3.16054 & 3.13140 \\
                       & $R_{\mathrm{circ}} [\mathrm{km}]$	 & 12.7302 & 12.7315 & 12.6439 & 12.6448 \\
    $\rrat=0.7$		   & $J [G\mathrm{M_\odot^2}/c]$	     & 3.58145 & 3.58078 & 3.57413 & 3.57317 \\
					   & $\Omega_c [\mathrm{rad \ s^{-1}}]$  & 8894.29 & 8892.15 & 9486.42 & 9486.16 \\
					   & $f_c [Hz]$				             & 1415.57 & 1415.23 & 1509.81 & 1509.77 \\
					   & $T/|W|$				             & 0.11547 & 0.11544 & 0.11516 & 0.11519 \\
    \hline
    \end{tabular}
\label{Tab:FlatStar_vs_Lorene_Hmax-04}
\end{table*}

The LORENE library also includes a tool to perform calculations of differentially rotating compact stars, RotDiff. In order to compare results of RotDiff and FlatStar we calculated sequences of differentially rotating strange stars for two fixed values of the maximum density (maximum enthalpy) $\rhm = 0.82 \cdot 10^{15}$ and $1.70 \cdot 10^{15}\,\mathrm{g/cm^3}$ ($\Hm = 0.2$ and $0.4$) as well as of the degree of differential rotation~$\Atil = 0.1$ and $0.2$. As will be explained in more detail in Section~\ref{sec:res}, each of those sequences starts from a non-rotating spherical configuration ($\Omega_c=0$ with $\rrat=r_p/r_e=1$) and terminates at the Keplerian limit (with $\Omega_c\ne 0$ and $0<\rrat<1$), when matter from the equator can no longer be retained by the gravitational attraction if the star spins faster. The results of these simulations are summarized on Figures~\ref{Fig:FlatStar_vs_Lorene_Hmax-02} and \ref{Fig:FlatStar_vs_Lorene_Hmax-04} which display the evolution of the gravitational mass $M$ as a function of $\rrat$ along such sequences.

As can be observed, it was not possible with LORENE to build complete sequences of configurations for the chosen values of the parameters. While the agreement with FlatStar is very good for slowly rotating stars, the code RotDiff cannot deal with very flattened stars, as was already observed for proto-neutron stars in \citet{VPCG}. More precisely, the lowest value of $\rrat$ that could be reached for strange stars with LORENE was $\sim 0.52$. On the other hand, FlatStar could follow sequences up to their physical endpoint, as is illustrated by Fig.~\ref{Fig:Shape_TypA} which depicts the meridian cross section of a SQS at the Keplerian limit with $\Atil = 0.2$ and $\rhm = 0.82 \cdot 10^{15} \mathrm{g/cm^3}$ ($\Hm = 0.2$). The detailed results of the comparison between FlatStar and LORENE for differentially rotating strange stars are gathered in Tabs.~\ref{Tab:FlatStar_vs_Lorene_Hmax-02} (for $\Hm = 0.2$) and \ref{Tab:FlatStar_vs_Lorene_Hmax-04} (for $\Hm = 0.4$), in which two values of $\rrat$ ($0.9$ and $0.7$) are taken into account.

\section{Differentially rotating Strange Quark Stars}
\label{sec:res}

\subsection{Building sequences}

Our first step in building sequences of configurations was to calculate the structure of spherically symmetric stars with given maximum densities $\rhm$. Then, we obtained a complete sequence by keeping a value of $\rhm$, fixing the degree of differential rotation $\Atil$ (with, in the case of rigid rotation, $\Atil=0$), and using a third parameter to spin up the star, e.g. by increasing the central angular velocity $\Omega_c$ or by decreasing the ratio of polar and equatorial radii $\rrat$. We remind the reader that the algorithmic of our code makes it possible to fix or vary any physical parameter \citep[see][for more detail]{AGV,GKVAK}.

\begin{figure}
\centering
  \includegraphics[width=0.65\linewidth, angle=-90]{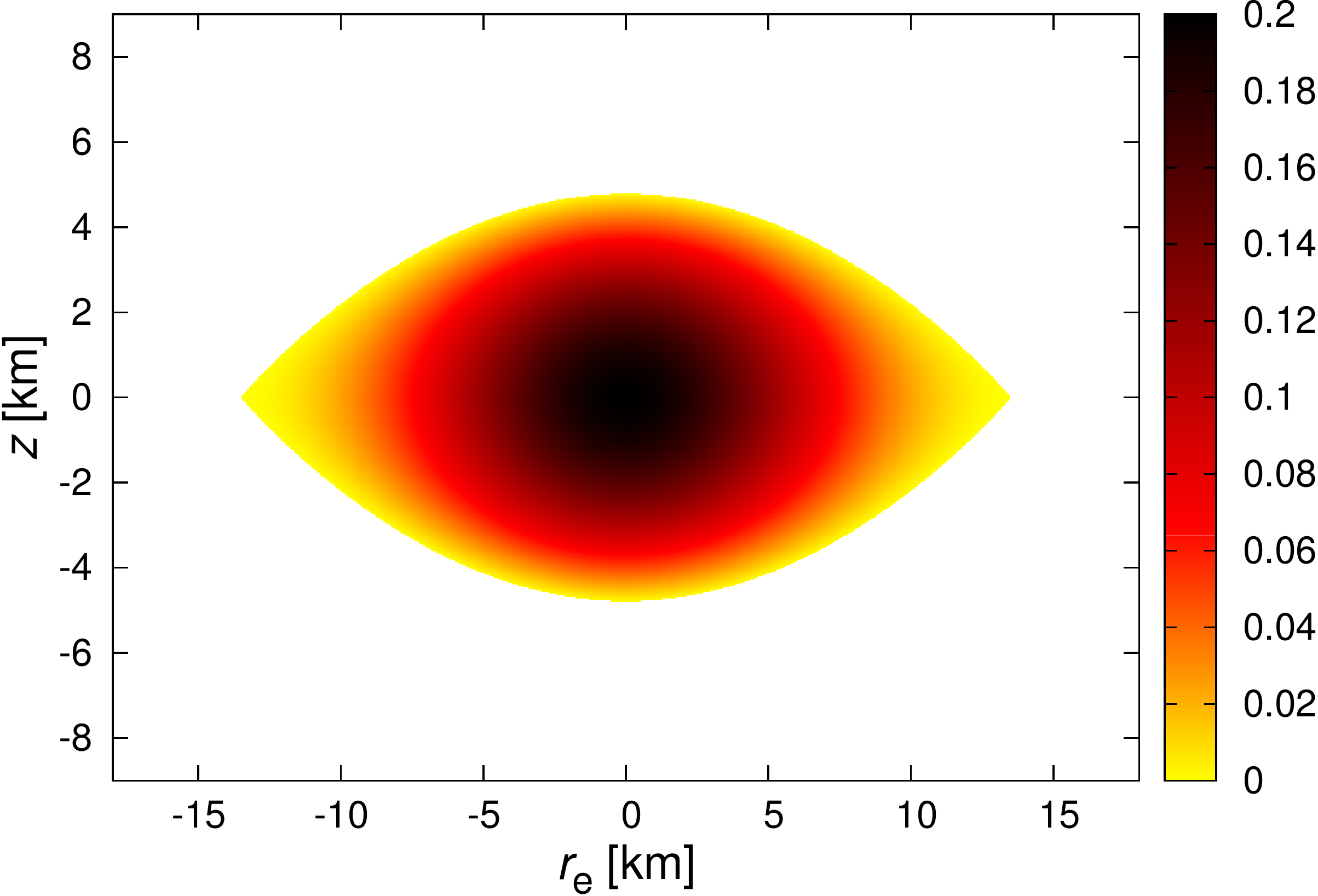}
  \caption{Isolines of enthalpy~$H$ in a meridian cross section of a Strange Quark Star at the mass-shedding limit with $\tilde{A} = 0.2$ and 
  $\rhm = 0.82 \cdot 10^{15}\,\mathrm{g/cm^3}$ ($\Hm = 0.2$). This configuration belongs to a type A sequence.}
  \label{Fig:Shape_TypA}
\end{figure}

As was explained in \citet{AGV} for homogeneous stars or $N=1$ polytropes and by \citet{SKGVA} for other adiabatic indices such sequences end in different way depending on the degree of differential rotation $\Atil$. Indeed, when the latter is sufficiently low, the sequences with fixed $\rhm$ terminate at the mass-shedding limit with a finite value of $\rrat$, as sequences of rigidly rotating stars do. As stated above, the fact that this limit is reached shows itself by the appearance of cusps along the equator of the star, giving it a shape similar to the one that can be seen on Fig.~\ref{Fig:Shape_TypA}. Such sequences are referred to as of type A. However, when $\Atil$ is large enough, a sequence with a fixed value of $\rhm$ can be extended down to $\rrat=0$, situation which indicates the entrance into the toroidal regime and at which we arbitrarily ended the sequence (as in our previous studies). In that case, the shape of the star looks like the one illustrated by Fig.~\ref{Fig:Shape_TypC} and the sequence is said to be of type C. More precisely, for each value of $\rhm$, there is a critical value $\Atc(\rhm)$ of $\Atil$ such that a sequence with fixed $\rhm$ and $\Atil$ starting from a non-rotating star ends at the mass-shedding limit (with a finite non-zero $\rrat$) if $\Atil < \Atc(\rhm)$ but enters into the toroidal regime (when $\rrat=0$) if $\Atil > \Atc(\rhm)$. This phenomenon, discovered for homogeneous stars and $N=1$ polytropes in \citet{AGV} as well as for polytropes with other indices in \citet{SKGVA}, was confirmed for SQS as we shall see below.

\begin{figure}
\centering
  \includegraphics[width=0.65\linewidth, angle=-90]{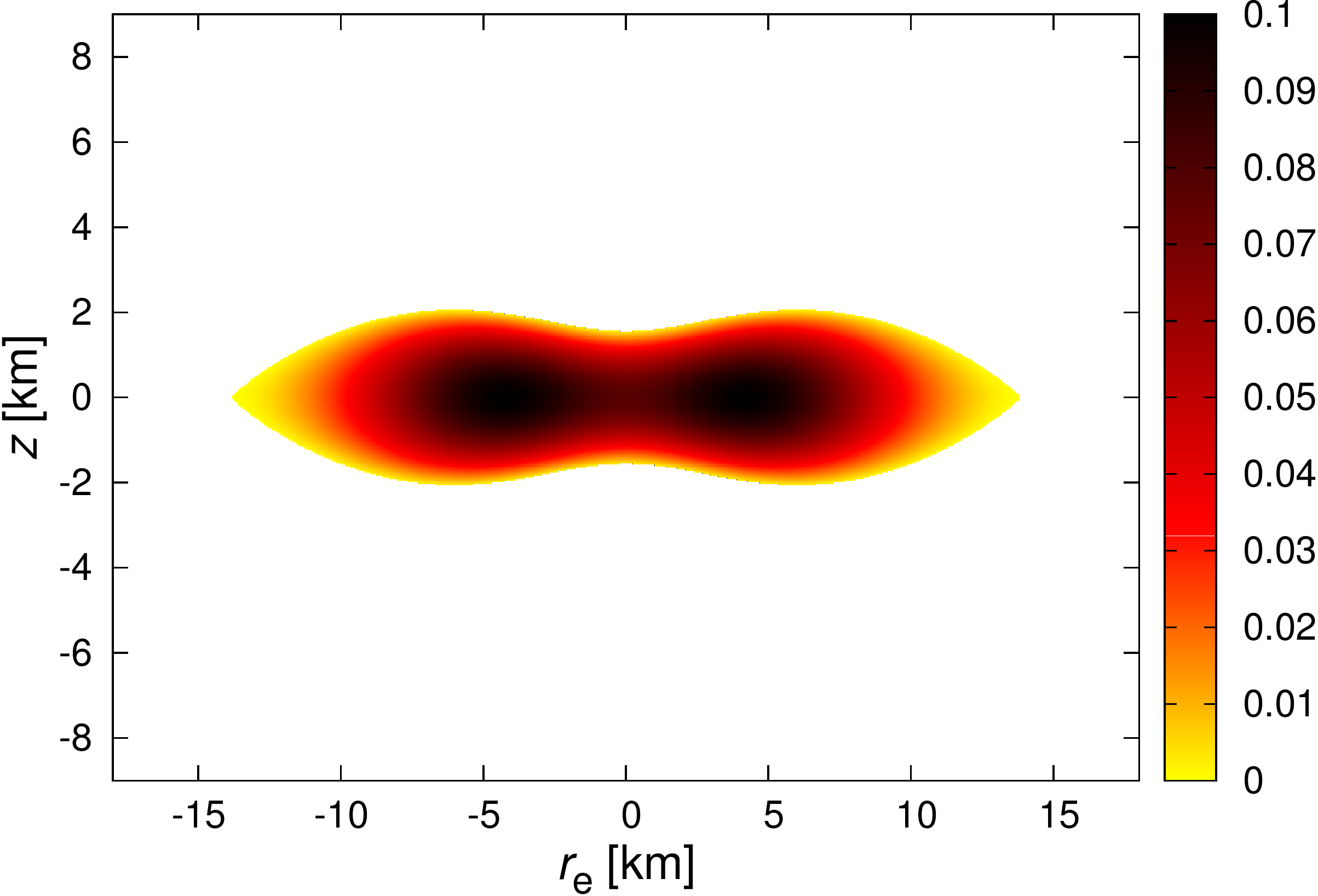}
  \caption{Isolines of enthalpy~$H$ in a meridian cross section of a Strange Quark Star at the mass-shedding limit with $\tilde{A} = 0.2$ and $\rhm = 0.59 \cdot 10^{15}\,\mathrm{g/cm^3}$ ($\Hm = 0.1$). This configuration belongs to a type B sequence.}
  \label{Fig:Shape_TypB}
\end{figure}

\begin{figure}
\centering
  \includegraphics[width=0.65\linewidth, angle=-90]{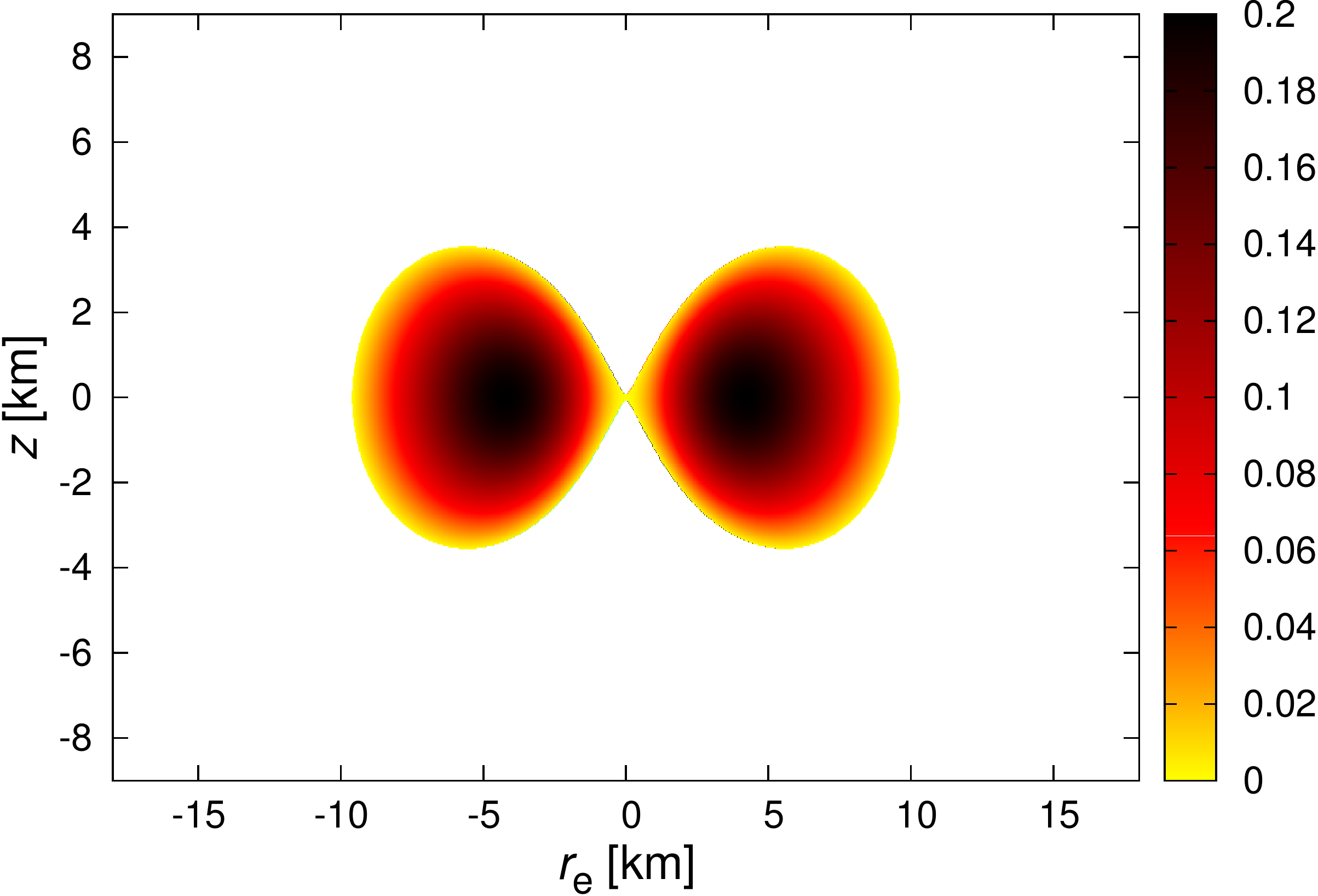}
  \caption{Isolines of enthalpy~$H$ in a meridian cross section of a Strange Quark Star close to the entrance into the toroidal regime ($\rrat=0.01$) with $\tilde{A} = 0.5$ and $\rhm = 0.82 \cdot 10^{15}\,\mathrm{g/cm^3}$ ($\Hm = 0.2$). This configuration belongs to a type C sequence.}
  \label{Fig:Shape_TypC}
\end{figure}

\begin{figure}
\centering
  \includegraphics[width=0.65\linewidth, angle=-90]{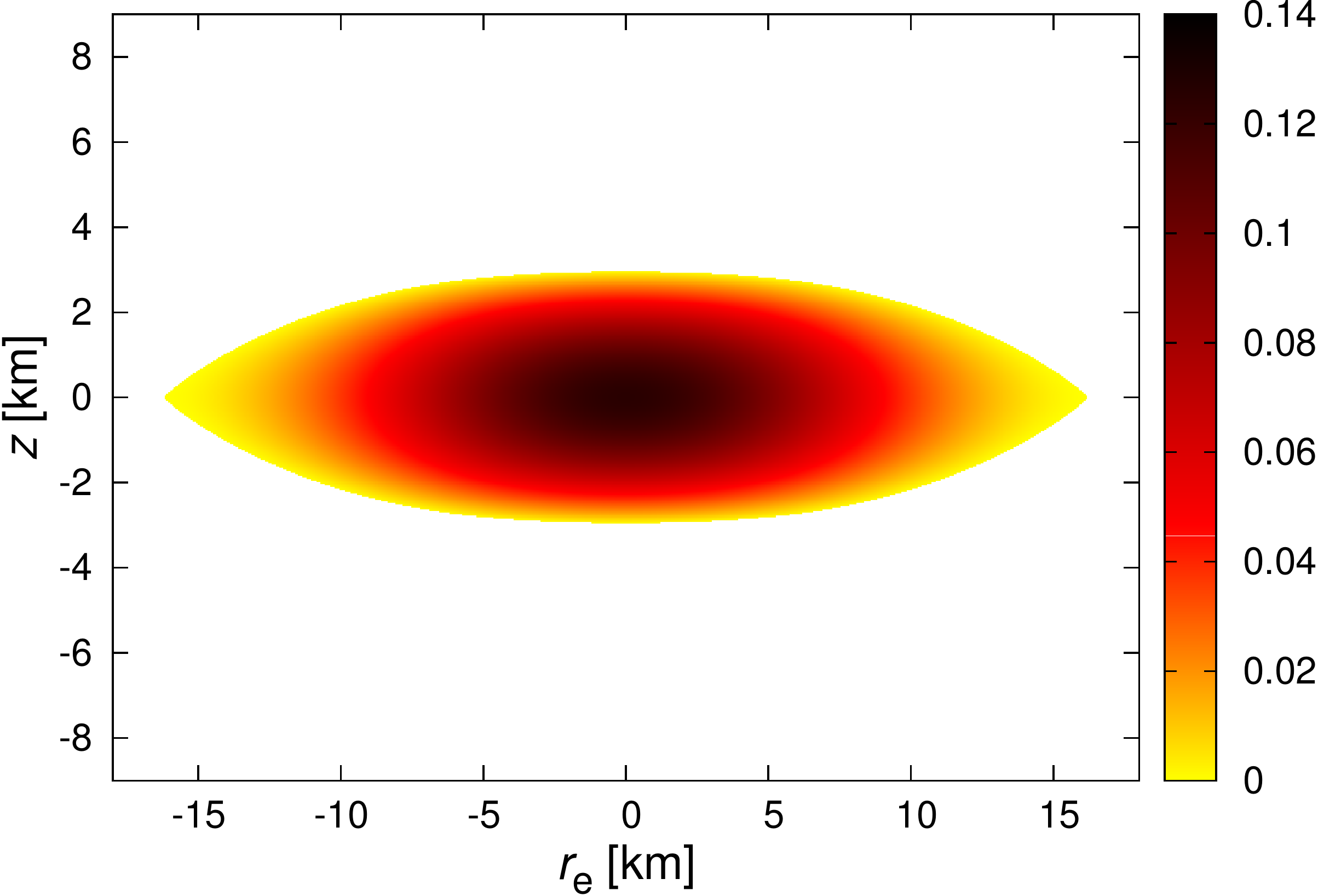}
  \caption{Isolines of enthalpy~$H$ in a meridian cross section of a Strange Quark Star at the mass-shedding limit with $\tilde{A} = 0.3$ and $\rhm = 0.64 \cdot 10^{15}\,\mathrm{g/cm^3}$ ($\Hm = 0.125$). This configuration belongs to a type D sequence.}
  \label{Fig:Shape_TypD}
\end{figure}

Before discussing in more detail the critical value of $\Atil$ which describes the transition between the A and C types of sequences, one shall remind the reader of another conclusion of \citet{AGV} and \citet{SKGVA}: the structure of the solution space of differentially rotating relativistic stars is even richer since, for $\Atil\sim\Atc(\rhm)$, it also contains some sequences of configurations  that, for fixed $\rhm$ and $\Atil$, do not have a non-rotating limit. Those configurations, called of types B and D, which coexist with type A and type C respectively. They are much more complex to obtain with a numerical code, which is why they were ignored before \citet{AGV}. For strange matter as well, we were able to numerically calculate their properties, and their typical shapes are illustrated by meridian cross sections on Figs.~\ref{Fig:Shape_TypB} (type B) and \ref{Fig:Shape_TypD} (type D).

\begin{figure}
\centering
  \includegraphics[width=0.7\linewidth, angle=-90]{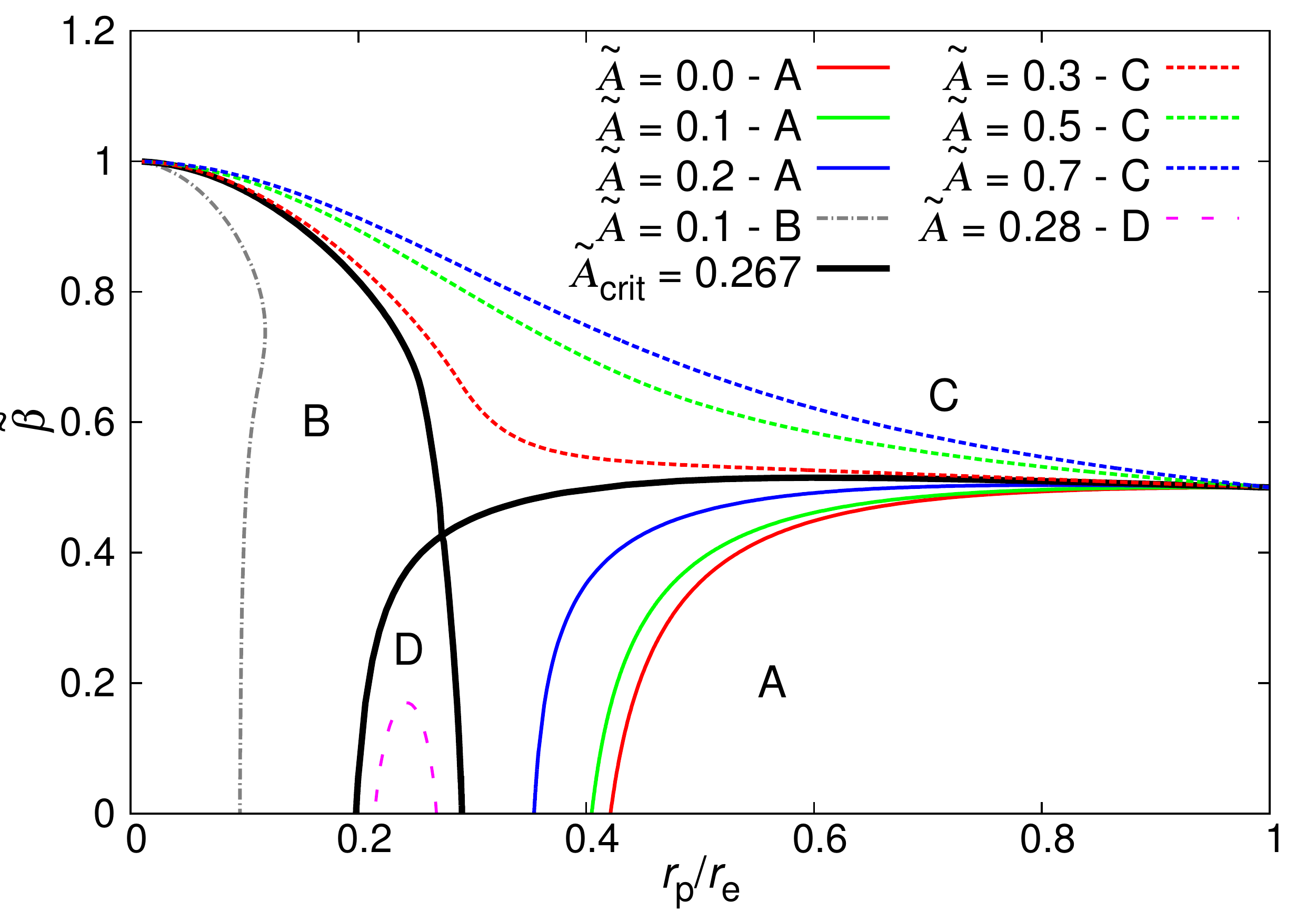}
    \caption{Illustration of the typical structure of the solution space showing four different types of sequences with given $\Atil$ in a plane with fixed maximal density $\rhm = 0.82 \cdot 10^{15}\,\mathrm{g/cm^3}$ (maximal enthalpy $\Hm = 0.2$). The axis are labelled by the normalized shedding parameter $\tilde{\beta}$ (with $\tilde{\beta}=0.5$ for a non-rotating star, $\tilde{\beta}=0$ at the mass-shedding limit and $\tilde{\beta}=1$ at the entrance into the toroidal regime) and the ratio between polar and equatorial radii $\rrat$. The bold line corresponds to the critical value of the degree of differential rotation $\Atc$, which divides the solution space into four regions associated to the A, B, C and D types. This figure, obtained for strange quark stars, is quite similar to the equivalent ones for polytropes and constant density stars [see \citet{AGV}, \citet{GKVAK} and \citet{SKGVA}]}
  \label{Fig:beta_vs_rr}
\end{figure}

By looking at those representative examples, it can be deduced that stars of those types are characterized by small values of $\rrat$. This feature and the global structure of the solution space for a given value of $\rhm$ are nevertheless better captured when one represents the sequences of configurations with fixed $\rhm$ and $\Atil$ in a $(\tilde{\beta}, \rrat)$ plane, as illustrated by Fig.~\ref{Fig:beta_vs_rr} for $\rhm=0.82 \cdot 10^{15}\,\mathrm{g/cm^3}$ ($\Hm=0.2$). On this figure, one easily sees that the plane is divided, by the curve made of configurations with the critical degree of differential rotation $\Atc(\rhm)$, in four domains corresponding to the four types of sequences. As was already stated in \citet{AGV}, \citet{GKVAK} and \citet{SKGVA}, type B sequences start at the mass-shedding limit $(\tilde{\beta}=0)$ and end at ($\tilde{\beta}=1, \rrat=0$), while type D sequences both start and end at the mass-shedding limit.

\begin{figure}
\centering
  \includegraphics[width=0.7\linewidth, angle=-90]{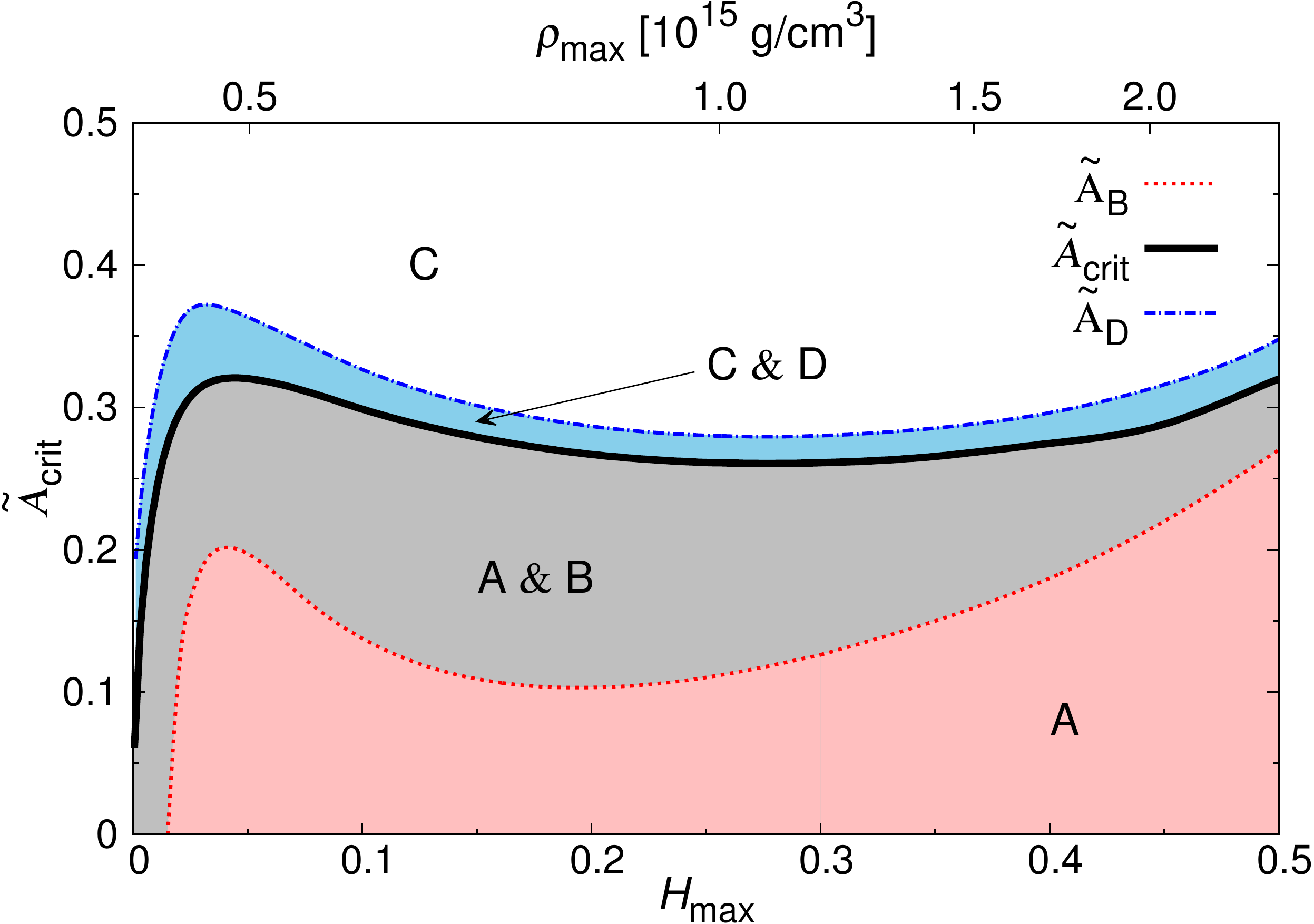}
    \caption{Critical values of the degree of differential rotation $\Atil$ as functions of the maximum enthalpy $\Hm$ (bottom axis) and of the maximum density $\rhm$ (top axis) for the MIT Bag model. The black line corresponds to $\Atc$, the blue line to the maximum value of $\Atil$ allowing type D, and the red line to the minimum value of $\Atil$ allowing type B (see text for more detail).}
  \label{Fig:separatrix_all}
\end{figure}

Another fact that can be observed on Fig.~\ref{Fig:beta_vs_rr} is that stars of types B and D exist in reduced ranges of degree of differential rotation (especially type D), type B occurring for $\Atil_B\leq\Atil\leq\Atc$ (always when type A exists as well but type C doesn't), and type D for $\Atil_D\geq\Atil\geq\Atc$ (always when type C exists as well but type A no longer does). This property is illustrated by Fig.~\ref{Fig:separatrix_all}. On the latter are plotted, for strange quark stars modeled by the MIT Bag model EOS~(\ref{Eq_of_state}), and as functions of $\rhm$ (top axis), or equivalently of $\Hm$ (bottom axis):
\begin{itemize}
\item as a black curve, the critical value $\Atc$;
\item as a blue curve, the value $\Atil_D$ such that for $\Atil_D\geq\Atil\geq\Atc$ types D and C coexist, but for $\Atil>\Atil_D$ only type C is left;
\item as a red curve, the value $\Atil_B$ such that for $\Atil_B\leq\Atil\leq\Atc$ types B and A coexist, but for $\Atil<\Atil_B$ only type A is left.
\end{itemize}

Comparing with the results of \citet{AGV} and of \citet{SKGVA}, it can be noticed on this figure that when $\rhm$ is sufficiently small ($\rhm \sim 0.5 \cdot 10^{15} \mathrm{g}/\mathrm{cm^3}$, equivalent to $\Hm \sim 0.05$, and below), the critical value attached to the MIT Bag model $\Atc$ is very close to the one for an incompressible fluid, but when $\rhm$ increases, the critical value for the compressible quark fluid becomes larger and reaches values close to those obtained for (polytropic) neutron stars [see Figs.~4, 6 of \citet{AGV} and of \citet{SKGVA}]. This behaviour is in agreement with the fact that for the MIT Bag model EOS~(\ref{Eq_of_state}) the adiabatic index $\gamma=d \ln P/d \ln n$, where $n$ is the baryon number density, diverges when $\rho\to\rho_0$ (low density limit) and has $a+1$ as a limit for $\rho\to\infty$ \citep[see][]{HPY}. Finally, the specific values of $\Atc$ also lead to the conclusion that for most $\rhm$, $\Atc(\rhm)$ is around $0.3$ so that sequences will generally be of type A (or B) for $\Atil<0.3$ and of type C for $\Atil>0.3$ (D being quite rare).

When looking for the maximum mass of a sequence of configurations with fixed $\rhm$ and $\Atil$, as we shall do in the following, having in mind the value of $\Atc(\rhm)$ and the type of the sequence which is followed is quite important. Indeed, while for a type A sequence (that ends at the mass-shedding limit and exists for $\Atil<\Atc$) the maximum mass on the sequence is a well-defined concept, it is not for those of types B and C which enter into the toroidal regime. As already explained, this difference is due to the fact that terminating type B and C sequences at $\rrat=0$ is an arbitrary choice and the mass could {\it a priori} become even larger for other non-simply-connected configurations with identical values of $\rhm$ and $\Atil$. In fact, as will be discussed in the next Subsection, the maximum mass of type C sequences was always found for $\rrat=0$, like for polytropic EOS \citet{GKVAK} and \citet{SKGVA} showing that it is more a supremum than an extremum.

\subsection{Maximum mass of strange quark stars}

As stated earlier, for a given EOS, a configuration of a differentially rotating star with the \citet{KEH} rotation law is prescribed by the values of three parameters. One usually sets the maximum density $\rhm$, which can be different from the central one $\rho_c$, the degree of differential rotation $\Atil$ and a third quantity that describes how fast the star rotates (e.g. $\Omega_\mathrm{c}$ the central angular velocity, $\rrat$ the ratio between the radii, or $J$ the angular momentum). As a consequence, the first step in the usual way of proceeding to determine the maximum mass of differentially rotating stars consists in building sequences with fixed $\Atil$ and $\rhm$, and using them to calculate, for various chosen values of $\Atil$, the highest mass as a function of $\rhm$. Finally, the maximum of those functions is the maximum mass as a function of $\Atil$, taking into account all $\rhm$ and all rates of rotation. The only subtlety of the described procedure is that, as explained earlier, the highly non-linear nature of the equations makes the solution space quite complicated with several types of configurations possibly coexisting, even for fixed $\Atil$, $\rhm$ and $\rrat$. This property is especially crucial to have in mind when one calculates configurations of stars close to the critical curve $\Atil=\Atc(\rhm)$ visible as a bold line in Fig.~\ref{Fig:beta_vs_rr}, situation in which an insufficiently robust numerical code can jump from one type of sequences to another. In the following, taking advantage of the high precision of the FlatStar code, we determine maximum masses for sequences  with fixed $\Atil$ and with a given type. 

\begin{figure}
\centering
  \includegraphics[width=0.7\linewidth, angle=-90]{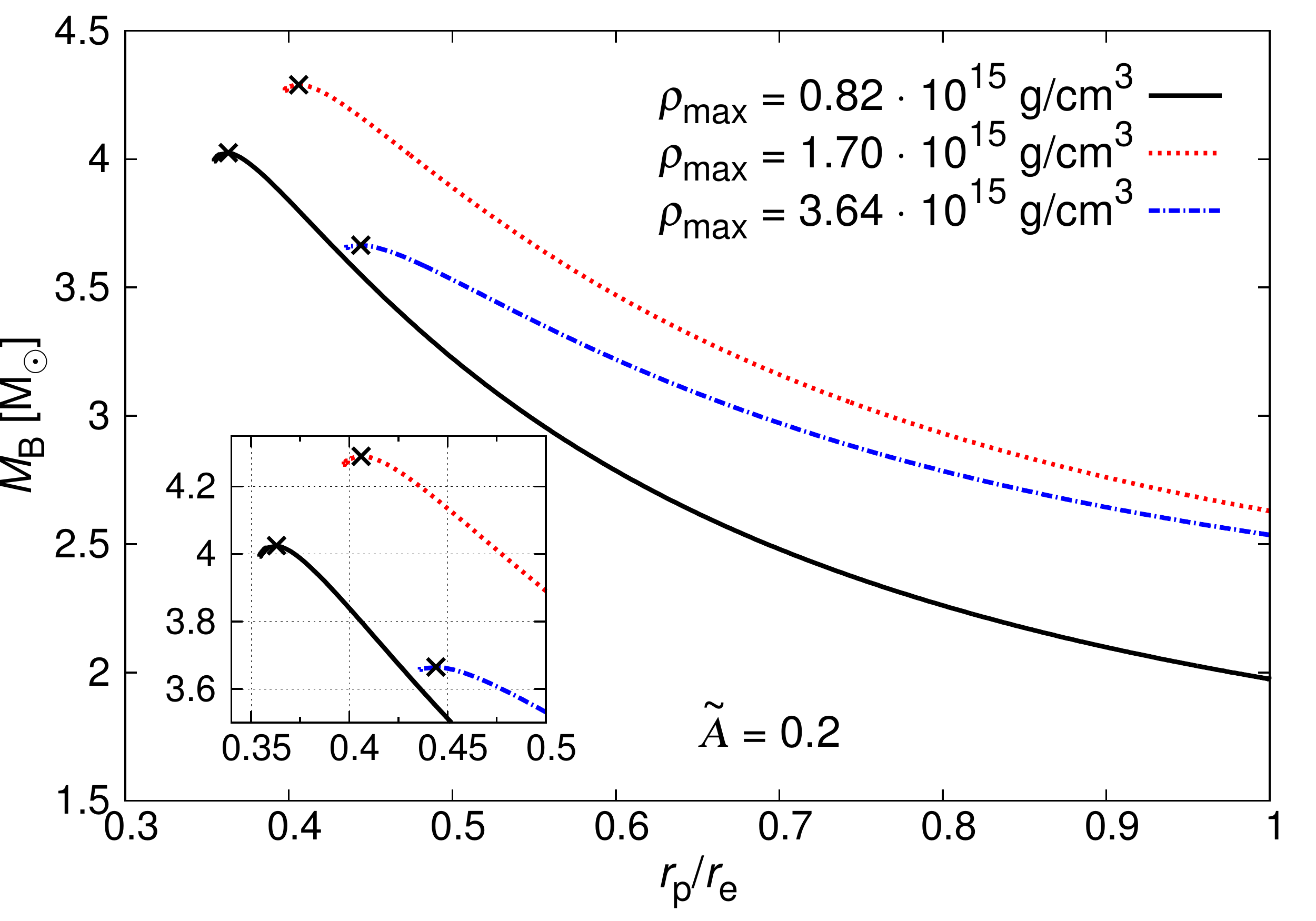}
     \caption{Baryonic mass $M_B$ as a function of $\rrat$ for three values of the maximum density (enthalpy) $\rhm = 0.82, 1.70, 3.64 \cdot 10^{15} \mathrm{g/cm^3}$ ($\Hm = 0.2, 0.4, 0.6$), with a fixed degree of differential rotation $\Atil = 0.2$. Black crosses correspond to the configurations with the highest masses. Since for the chosen $\rhm$, $\Atc(\rhm)>0.2$, those sequences, which start from non-rotating configurations, are of type A and consequently end at the Keplerian limit.}
  \label{Fig:M_rr_TypeA}
\end{figure}

Applying the preceding process, we represented on Figure \ref{Fig:M_rr_TypeA} the baryonic mass $M_B$ as a function of $\rrat$ for three complete evolutionary sequences with the same fixed degree of differential rotation $\Atil = 0.2$ but different values of the maximum density (enthalpy), $\rhm = 0.82, 1.70, 3.64 \cdot 10^{15} \mathrm{g/cm^3}$ (equivalently $\Hm = 0.2, 0.4, 0.6$). For such $\Atil$, Fig.~\ref{Fig:separatrix_all} shows that the sequences are of type A and end at the mass-shedding limit. As can be seen on Fig.~\ref{Fig:M_rr_TypeA}, the highest values of the mass are obtained for configurations (marked with a cross) close to but not at the mass-shedding limit, as was already observed for polytropes in \citet{GKVAK}. Notice that the star represented on Fig.~\ref{Fig:Shape_TypA} lies at the end (at the mass-shedding limit) of the sequence plotted with a black line on Fig~\ref{Fig:M_rr_TypeA}.

\begin{figure}
\centering
  \includegraphics[width=0.7\linewidth, angle=-90]{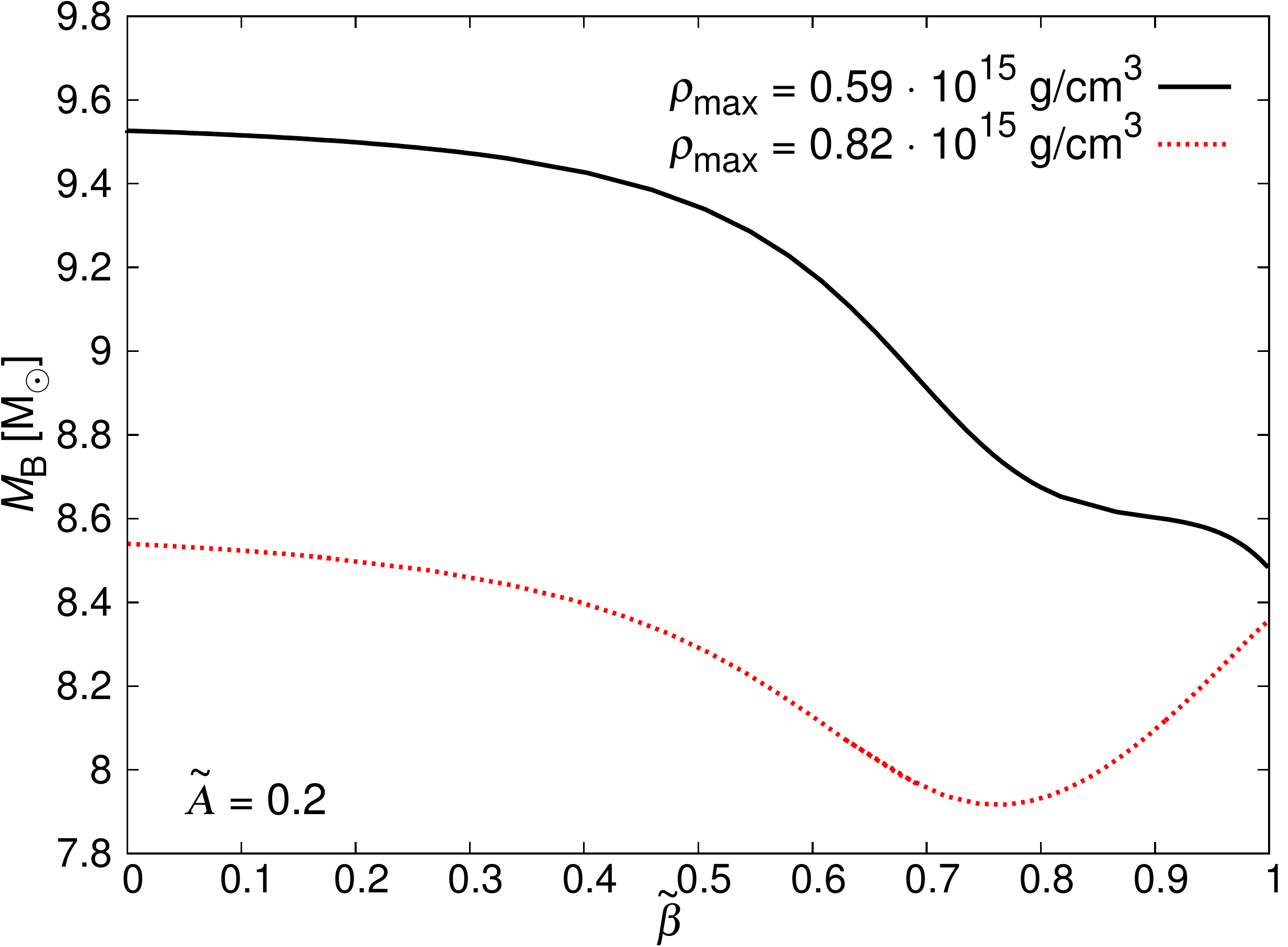}
     \caption{Baryonic mass $M_B$ as a function of $\tilde{\beta}$ for two values of the maximum density (enthalpy) $\rhm = 0.59$ and $0.82 \cdot 10^{15} \mathrm{g/cm^3}$ ($\Hm = 0.1$ and $0.2$), with a fixed degree of differential rotation $\Atil = 0.2$. For the chosen $\rhm$, $\Atil_B\leq0.2\leq\Atc$, so that the sequences can be of type B, which they are. Consequently they both start at the mass-shedding limit ($\tilde{\beta}=0$) and end at the entrance into the toroidal regime, $\rrat \rightarrow 0$ and $\tilde{\beta}=1$. For all such sequences, the highest mass is found at the mass-shedding limit which also corresponds to the highest value of $\rrat$.}
  \label{Fig:M_rr_TypeB}
\end{figure}

\begin{figure}
\centering
  \includegraphics[width=0.7\linewidth, angle=-90]{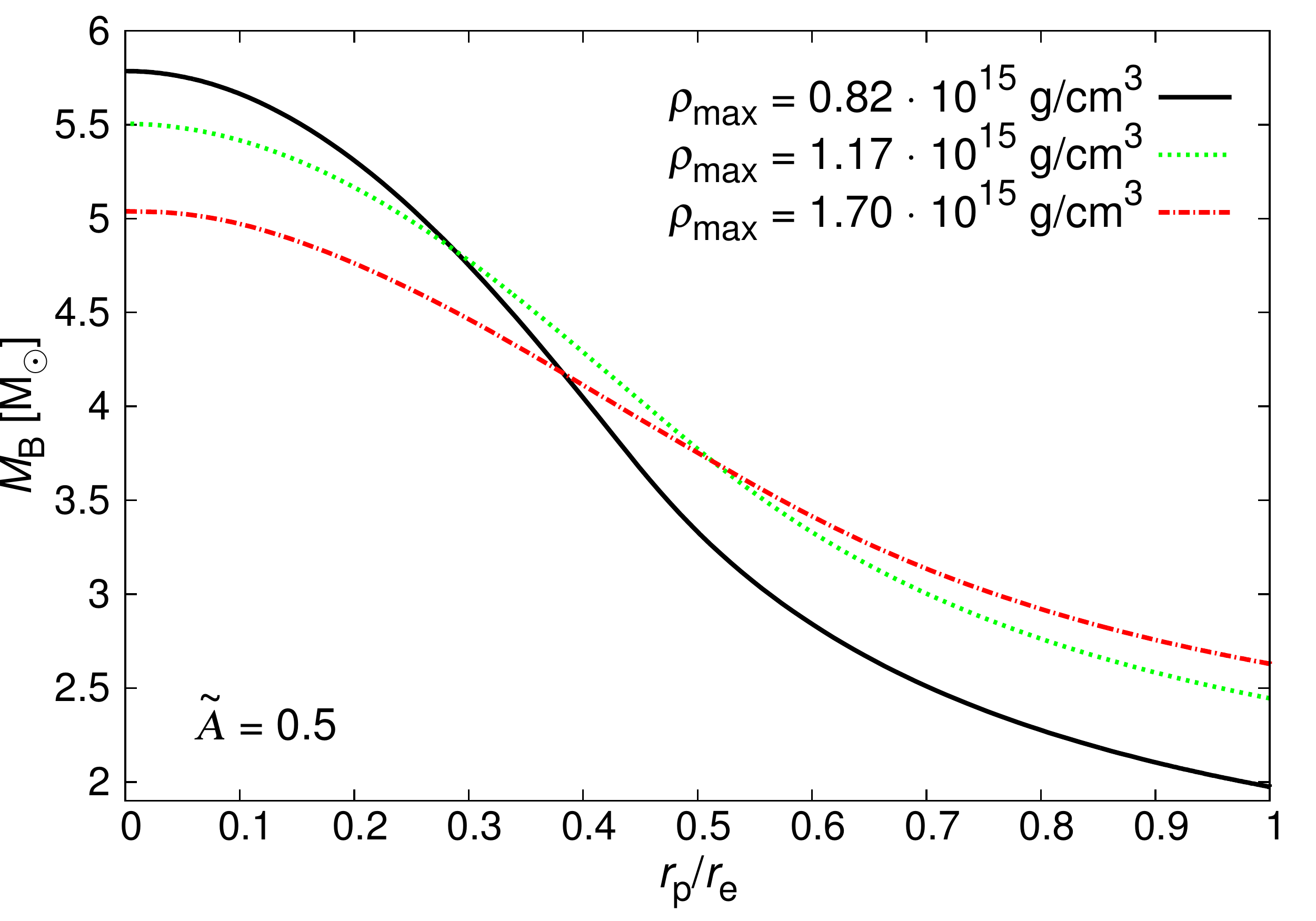}
     \caption{Baryonic mass $M_B$ as a function of $\rrat$ for three values of the maximum density (enthalpy) $\rhm = 0.82, 1.17, 1.70 \cdot 10^{15} \mathrm{g/cm^3}$ ($\Hm = 0.2, 0.3, 0.4$), with a fixed degree of differential rotation $\Atil = 0.5$. Since for the chosen $\rhm$, $\Atc(\rhm)<0.5$, those sequences, which start from non-rotating configurations, are of type C and consequently end at the entrance in the toroidal regime, $\rrat \rightarrow 0$. For all of them, the highest mass is found at this specific point.}
  \label{Fig:M_rr_TypeC}
\end{figure}

On Fig.~\ref{Fig:M_rr_TypeC}, the baryonic mass $M_\mathrm{B}$ is displayed as a function of $\rrat$ for three complete evolutionary sequences with the same maximum density as on Fig.~\ref{Fig:M_rr_TypeA}, but this time with $\Atil = 0.5$ which is larger than $\Atc$ for those values of $\rhm$. Consequently, all sequences are of type C and reach the entrance into the toroidal regime at $\rrat=0$. As stated in the previous Section, one easily observes that on all those sequences the highest mass is at $\rrat=0$. Notice that the configuration from Fig.~\ref{Fig:Shape_TypC} lies at the end of the sequence plotted with a black line on Fig~\ref{Fig:M_rr_TypeC}.

\begin{figure}
\centering
  \includegraphics[width=0.7\linewidth, angle=-90]{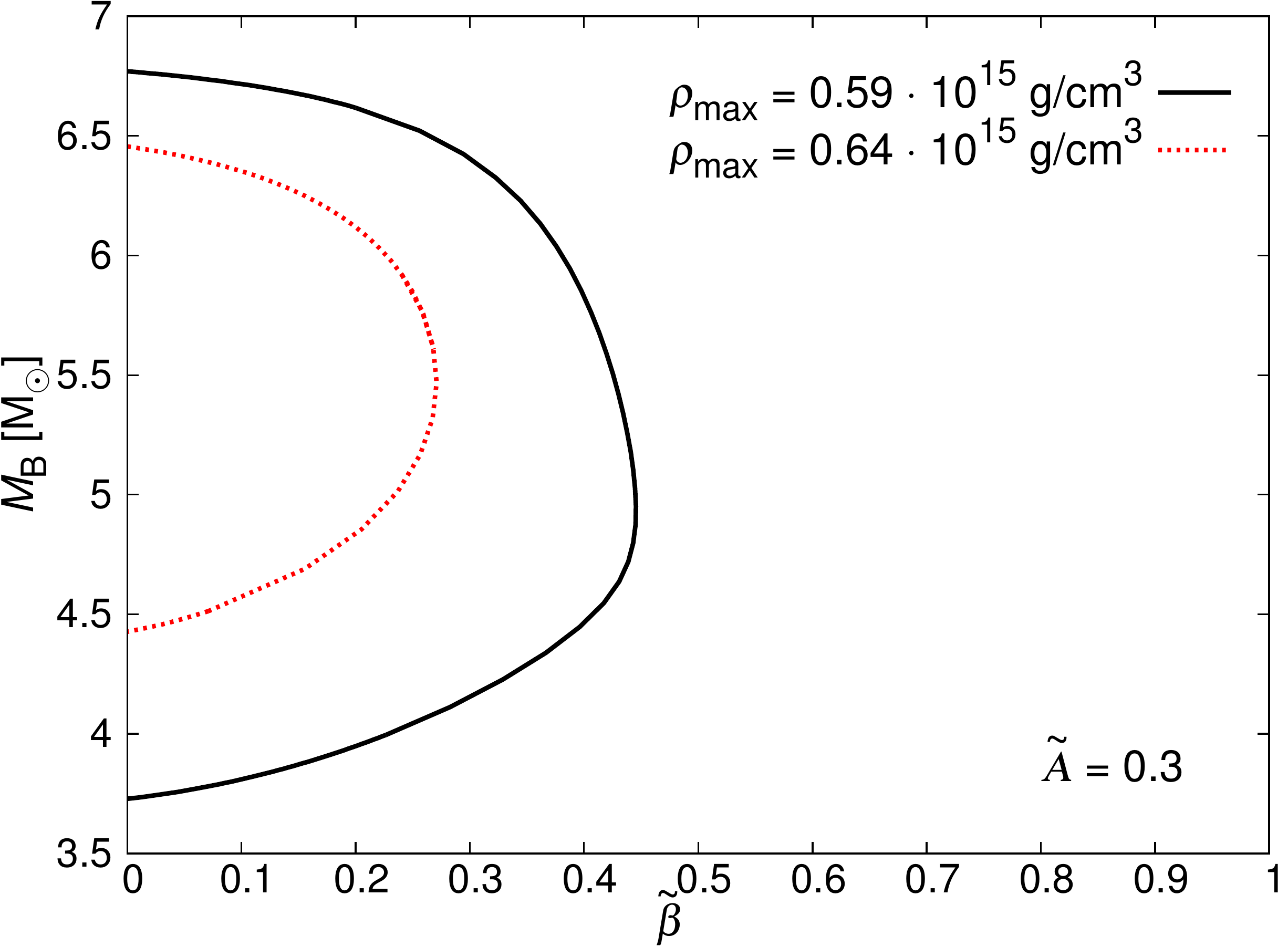}
     \caption{Baryonic mass $M_B$ as a function of $\tilde{\beta}$ for two values of the maximum density (enthalpy) $\rhm = 0.59$ and $0.64 \cdot 10^{15} \mathrm{g/cm^3}$ ($\Hm = 0.1$ and $0.125$), with a fixed degree of differential rotation $\Atil = 0.3$. Since for the chosen $\rhm$, $\Atil_D\geq0.3\geq\Atc$, the sequences can be of type D, which they are. Consequently they both start and end at the mass-shedding limit ($\tilde{\beta}=0$). In practice, we observed that for stars of this type, the highest mass was always found at the mass-shedding limit with the smallest value of $\rrat$.}
  \label{Fig:M_rr_TypeD}
\end{figure}

As already explained, configurations of types B and D do not have any direct connection with a non-rotating configuration, which means that they exist only for $\rrat<1.0$. On Fig.~\ref{Fig:M_rr_TypeB}, the baryonic mass $M_B$ is displayed as a function of the shedding parameter $\tilde{\beta}$ for two complete sequences with the same fixed $\Atil=0.2$ but different maximum density $\rhm = 0.59$ and $0.82 \cdot 10^{15} \mathrm{g/cm^3}$ (maximum enthalpy $\Hm = 0.1$ and $0.2$). Both are of type B, as can be realized by the fact that their extremities are at the mass-shedding limit ($\tilde{\beta}=0$) and at the entrance into the toroidal regime  ($\tilde{\beta}=1$ and $\rrat=0$). We observed that sequences of this type have their highest mass at the Keplerian limit. For instance, the configuration depicted on Fig.~\ref{Fig:Shape_TypB} is the one with the highest mass of the sequence plotted as a black line on Fig.~\ref{Fig:M_rr_TypeB}.
 
Figure~\ref{Fig:M_rr_TypeD} shows curves equivalent to those of Fig.~\ref{Fig:M_rr_TypeB}, but for two complete sequences of type D. They were calculated with the same fixed $\Atil=0.3$, but with different maximum density $\rhm = 0.59$ and $0.64 \cdot 10^{15} \mathrm{g/cm^3}$ (maximum enthalpy $\Hm = 0.1$ and $0.125$). Type D sequences start and end at the mass-shedding limit ($\tilde{\beta}=0$), and we observed, as was already the case for other EOSs, that the highest mass is found for the lowest value of $\rrat$. The meridian cross-section shown on Fig.~\ref{Fig:Shape_TypD} corresponds to the maximum mass configuration of the sequence plotted as a red line on Fig.~\ref{Fig:M_rr_TypeD}.
 
\begin{figure}
\centering
  \includegraphics[width=0.7\linewidth, angle=-90]{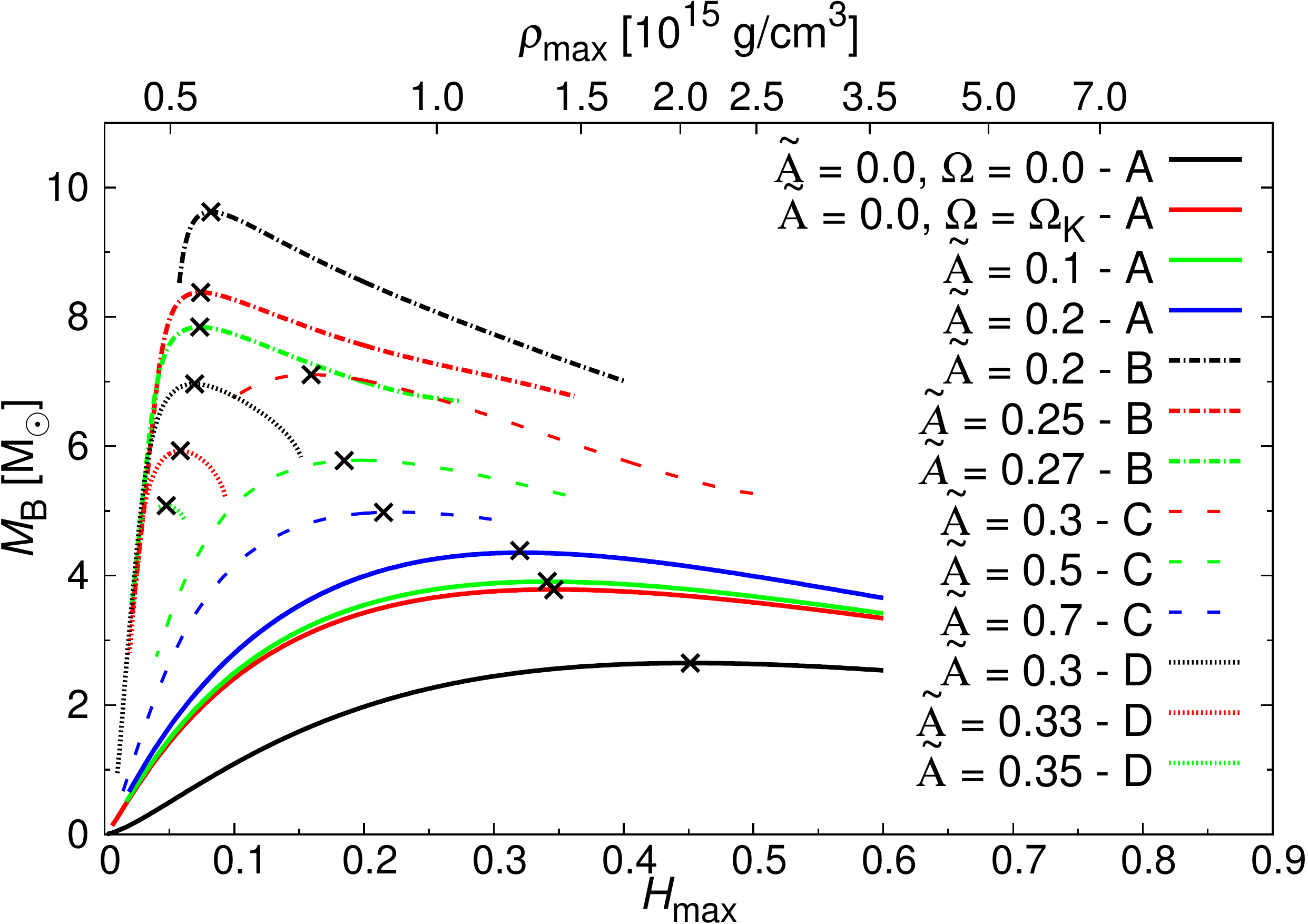}
  \caption{Highest baryonic mass $M_\mathrm{B}$ as a function of the maximum enthalpy $\Hm$ (bottom axis) or the maximum density $\rhm$ (top axis) for fixed values of the degree of differential rotation $\Atil$. The black solid line represents non-rotating stars, the red solid line rigid rotation (for which the highest mass is at the mass-shedding limit), while other lines correspond to differentially rotating stars. Solid lines stand for type A sequences for which the highest mass is reached close to the mass-shedding limit, dashed-line are for type C sequences on which the highest mass is obtained at $\rrat=0$, dashed-dotted lines are for type B sequences for which the highest mass is reached at the mass-shedding limit and finally dotted lines are for type D sequences whose highest mass is found at the mass-shedding limit for their smallest value of $\rrat$. For all values of $\Atil$, configurations with maximum mass are marked with a black cross.}
  \label{Fig:M_hmax}
\end{figure}

Doing similar calculations for broad ranges of $\rhm$, one eventually gets curves as those gathered on Fig.~\ref{Fig:M_hmax} which displays the results for various values of $\Atil$ with the convention that dotted lines are for type D sequences, dashed lines for type C, dashed-dotted lines for type B and solid lines for type A (for comparison, we plotted as well the curves associated to static spherical stars -- solid black line -- and to rigidly rotating stars -- solid red line--, for which the highest mass is obtained at the mass-shedding limit $\Omega=\Omega_K$).

\begin{table*}
\centering
    \caption{Properties of configurations with maximum mass of non-rotating ($\tilde{A} = 0.0$, $\Omega = 0.0$), rigidly rotating 
    ($\tilde{A} = 0.0$ - Keplerian limit) and differentially rotating ($\tilde{A} = 0.1, 0.2, 0.3, 0.5, 0.7$) Strange Quark Stars of types A and C. In the Table are given the baryonic mass $M_\mathrm{B}$, the gravitational mass $M$, the central and equatorial angular velocities $\Omega_c$ and $\Omega_\mathrm{eq}$, the corresponding frequencies $f_c$ and $f_\mathrm{eq}$, the central and maximal densities $\rho_c$ and $\rhm$, the corresponding enthalpies $H_c$ and $\Hm$, the circumferential radius $R_\mathrm{circ}$, the coordinate equatorial radius $r_e$, the ratio between the polar and equatorial radii $\rrat=r_p/r_e$, the angular momentum $J$, and the ratio between the kinetic and potential energies $T/|W|$.}
    \begin{tabular}{@{}llclllclll@{}}
    \hline
    \hline
     & $\tilde{A} = 0.0$ &  & $\tilde{A} = 0.0$ & $\tilde{A} = 0.1$ & $\tilde{A} = 0.2$ &  & $\tilde{A} = 0.3$ & $\tilde{A} = 0.5$ & $\tilde{A} = 0.7$ \\
     & $\Omega = 0.0$    &  & Keplerian limit   & \multicolumn{2}{c}{Type A}            &  & \multicolumn{3}{c}{Type C}                                \\
    \hline
    $M [\mathrm{M_\odot}]$		            & 1.96372 &  & 2.81847 & 2.91624 & 3.26356 &  & 5.25630 & 4.24179 & 3.67839 \\
    ${M_{\mathrm{B}} [\mathrm{M_\odot}]}$                 		    & 2.64888 &  & 3.78386 & 3.90858 & 4.38617 &  & 7.25072 & 5.77867 & 4.97946 \\
    $\Omega_{\mathrm{c}} [\mathrm{rad \ s^{-1}}]$ 	    & 0        &  & 9655.09 & 10066.3 & 11037.5 &  & 13703.2 & 19512.4 & 26019.3 \\
    $\Omega_{\mathrm{eq}} [\mathrm{rad \ s^{-1}}]$ 	    & 0        &  & 9655.09 & 9561.17 & 9039.06 &  & 7669.96 & 7712.25 & 7555.35 \\
    $f_{\mathrm{c}} [Hz]$ 	    		  	            & 0        &  & 1536.66 & 1602.09 & 1756.67 &  & 2180.93 & 3105.50 & 4141.09 \\
    $f_{\mathrm{eq}} [Hz]$ 	    		 	            & 0        &  & 1536.66 & 1521.71 & 1438.61 &  & 1220.71 & 1227.44 & 1202.47 \\
    $\rho_{\mathrm{c}} [\mathrm{10^{15} g \ cm^{-3}}]$  & 2.059    &  & 1.323    & 1.362    & 1.261    &  & 0.428   & 0.428   & 0.428 \\
    $\rho_{\mathrm{max}} [\mathrm{10^{15} g \ cm^{-3}}]$& 2.059    &  & 1.323    & 1.362    & 1.261    &  & 0.690    & 0.779    & 0.866 \\
    $H_{\mathrm{c}}$ 				    	            & 0.451396 &  & 0.33308  & 0.34105 & 0.32008 &  & 0.00043 & 0.00033 & 0.00023 \\
    $H_{\mathrm{max}}$ 				    	            & 0.451396 &  & 0.33308  & 0.34105 & 0.32008 &  & 0.14913 & 0.18473 & 0.21510 \\
    $R_{\mathrm{circ}} [\mathrm{km}]$ 		    	    & 10.7104 &  & 16.3338 & 16.5381 & 17.7295 &  & 21.9281 & 18.5692 & 16.6423 \\
    $r_e [\mathrm{km}]$ 	    		                & 7.53167 &  & 11.17434 & 11.1727 & 11.5620 &  & 10.6879 & 9.97624 & 9.46258 \\
    $\rrat$     		                                & 1        &  & 0.46792 & 0.45191 & 0.38979 &  & 0.01     & 0.01     & 0.01     \\
    $J [G\mathrm{M_\odot^2}/c]$ 	    		        & 0        &  & 6.96846 & 7.59663 & 10.1705 &  & 28.1603 & 17.4939 & 12.4682 \\
    $T/|W|$ 				    		                & 0        &  & 0.20721 & 0.21652 & 0.25269 &  & 0.33465 & 0.30538 & 0.27652 \\
    \hline
    \end{tabular}
\label{Tab:Mmax_AC}
\end{table*}

\begin{table*}
\centering
    \caption{Properties of configurations with maximum mass of differentially rotating ($\tilde{A}$ = 0.015, 0.2, 0.23, 0.25, 0.27, 0.29, 0.3, 0.33, 0.35) Strange Quark Stars of types B and D. In the Table are given the baryonic mass $M_\mathrm{B}$, the gravitational mass $M$, the central and equatorial angular velocities $\Omega_c$ and $\Omega_\mathrm{eq}$, the corresponding frequencies $f_c$ and $f_\mathrm{eq}$, the central and maximal densities $\rho_c$ and $\rhm$, the corresponding enthalpies $H_\mathrm{c}$ and $\Hm$, the circumferential radius $R_\mathrm{circ}$, the coordinate equatorial radius $r_e$, the ratio between the polar and equatorial radii $\rrat=r_p/r_e$, the angular momentum $J$, and the ratio between the kinetic and potential energies $T/|W|$.}
    \begin{tabular}{@{}llllllllclll@{}}
    \hline
    \hline
     & $\tilde{A} = 0.015$ & $\tilde{A} = 0.2$ & $\tilde{A} = 0.23$ & $\tilde{A} = 0.25$ & $\tilde{A} = 0.27$ & $\tilde{A} = 0.29$ &  & $\tilde{A} = 0.3$ & $\tilde{A} = 0.33$ & $\tilde{A} = 0.35$ \\
     & \multicolumn{6}{c}{Type B}                                                                                                  &  & \multicolumn{3}{c}{Type D}                                  \\
    \hline
    ${M [\mathrm{M_\odot}]}$    		                    & 7.59282 & 6.92478 & 6.48297 & 6.16558 & 5.83180 & 5.47377 &  & 5.27581 & 4.59474 & 4.01156 \\
    $M_{\mathrm{B}} [\mathrm{M_\odot}]$ 		    	& 10.8074 & 9.62320 & 8.88792 & 8.37731 & 7.84149 & 7.26888 &  & 6.96280 & 5.93151 & 5.08301 \\
    $\Omega_{\mathrm{c}} [\mathrm{rad \ s^{-1}}]$ 	    & 9447.75 & 9271.92 & 9181.15 & 9054.78 & 8882.55 & 8634.43 &  & 8500.81 & 7937.89 & 7427.45 \\
    $\Omega_{\mathrm{eq}} [\mathrm{rad \ s^{-1}}]$ 	    & 6823.31 & 6443.00 & 6269.71 & 6144.96 & 6014.84 & 5865.32 &  & 5791.96 & 5516.91 & 5286.19 \\
    $f_{\mathrm{c}} [Hz]$ 	    		  	            & 1503.66 & 1475.67 & 1461.23 & 1441.11 & 1413.70 & 1374.21 &  & 1352.95 & 1263.36 & 1182.12 \\
    $f_{\mathrm{eq}} [Hz]$ 	    		 	            & 1085.964 & 1025.44 & 997.855 & 978.001 & 957.292 & 933.495 &  & 921.818 & 878.043 & 841.323 \\
    $\rho_{\mathrm{c}} [\mathrm{10^{15} g \ cm^{-3}}]$  & 0.462    & 0.510    & 0.529    & 0.535    & 0.537    & 0.534    &  & 0.531    & 0.512    & 0.495    \\
    $\rho_{\mathrm{max}} [\mathrm{10^{15} g \ cm^{-3}}]$& 0.588    & 0.552    & 0.543    & 0.539    & 0.537    & 0.534    &  & 0.531    & 0.512    & 0.495    \\
    $H_{\mathrm{c}}$ 				    	            & 0.025510 & 0.05689 & 0.06837 & 0.07224 & 0.07340 & 0.07120 &  & 0.06940 & 0.05850 & 0.04760 \\
    $H_{\mathrm{max}}$ 				    	            & 0.10100 & 0.08200 & 0.07650 & 0.07409 & 0.07340 & 0.07120 &  & 0.06940 & 0.05850 & 0.04760 \\
    $R_{\mathrm{circ}} [\mathrm{km}]$ 		    	    & 29.9610 & 30.0117 & 29.8564 & 29.7765 & 29.6938 & 29.6490 &  & 29.5937 & 29.4891 & 29.3296 \\
    $r_e [\mathrm{km}]$ 	    		                & 12.9534 & 15.1022 & 16.2005 & 16.9673 & 17.7549 & 18.6191 &  & 19.0441 & 20.5223 & 21.6422 \\
    $\rrat$     		                                & 0.06203 & 0.09897 & 0.11268 & 0.11949 & 0.12494 & 0.12848 &  & 0.12997 & 0.12994 & 0.12594 \\
    $J [G\mathrm{M_\odot^2}/c]$ 	    		        & 60.6172 & 51.7050 & 46.0668 & 42.2461 & 38.3987 & 34.5055 &  & 32.4296 & 25.8446 & 20.7697 \\
    $T/|W|$ 				    		                & 0.36215 & 0.36655 & 0.36815 & 0.36915 & 0.37021 & 0.37174 &  & 0.37252 & 0.37675 & 0.38172 \\
    \hline
    \end{tabular}
\label{Tab:Mmax_BD}
\end{table*}

In fact, in order to calculate with a good accuracy the maximum masses (and other physical properties of the configurations with maximum mass), we proceeded to a more meticulous exploration of the solution space than what could suggest Figs.~\ref{Fig:M_rr_TypeA}, \ref{Fig:M_rr_TypeB}, \ref{Fig:M_rr_TypeC} and \ref{Fig:M_rr_TypeD}. For instance, in the case of type A sequences, we explored $(\Hm, \rrat)$ planes (for fixed $\Atil$) with a fine mesh, looking for the configuration with maximum mass using a 2D code, while for type C sequences it was sufficient to search for it on the $\rrat=0$ line (and to save some computational time without decreasing the precision of the results we could even end the sequences at $\rrat \sim 0.01$). See the Appendix of \citet{GKVAK} for more details on the method.

\begin{figure}
\centering
  \includegraphics[width=0.7\linewidth, angle=-90]{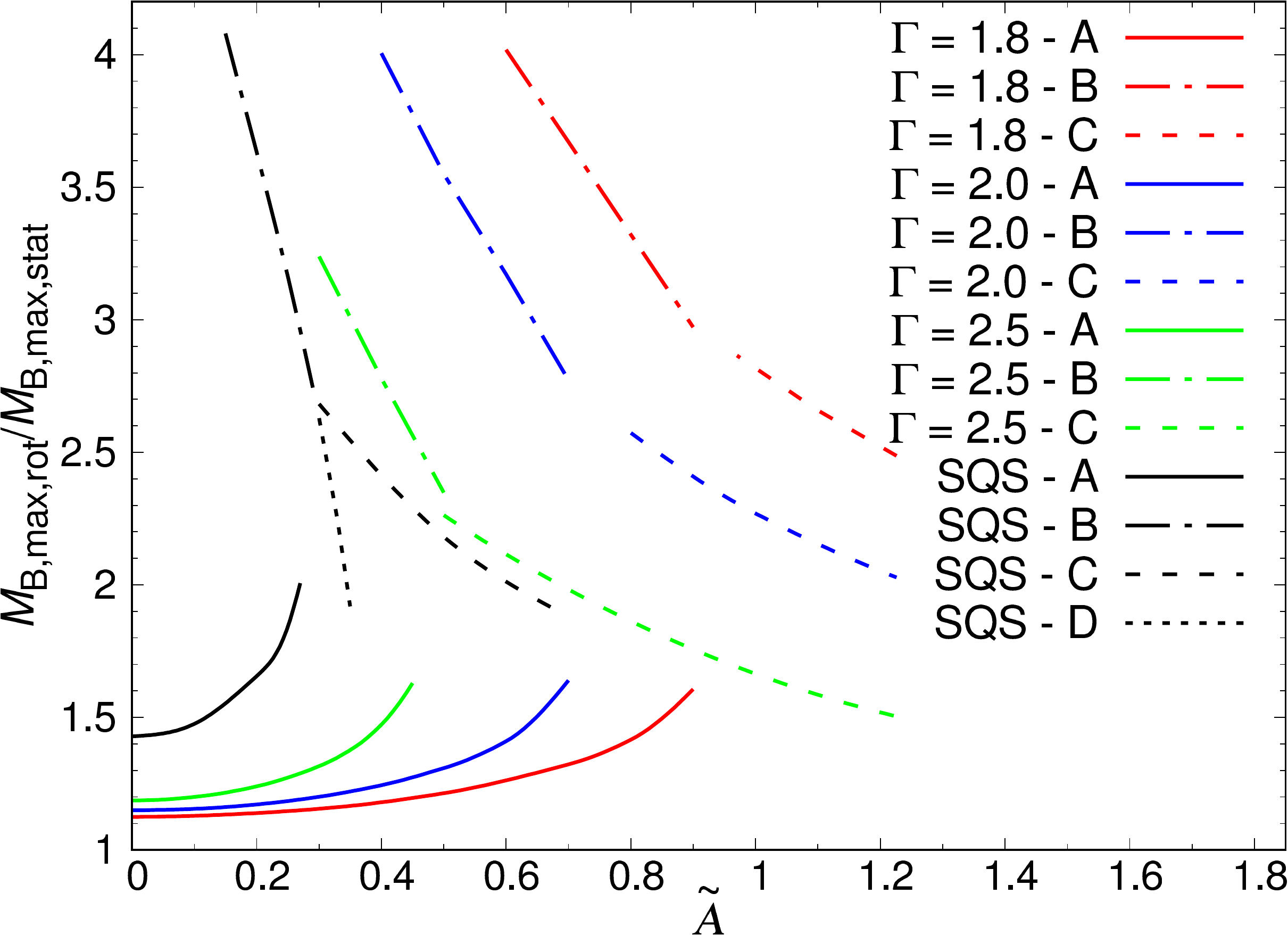}
  \caption{Maximum baryonic mass of differentially rotating strange quark stars and neutron stars described by a polytropic EOS normalized by the maximum mass of a non-rotating configuration with the same EOS as a function of the degree of differential rotation $\tilde{A}$. 
  Black lines correspond to our results for four types of differentially rotating SQS, other lines corresponds to differentially rotating polytropes: blue - $\Gamma = 2.0$ taken from \citet{GKVAK}, red - $\Gamma = 1.8$ and green - $\Gamma = 2.5$ respectively taken from \citet{SKGVA}.}
  \label{Fig:SQS_NSPoly_Comparison}
\end{figure}

\begin{figure}
\centering
  \includegraphics[width=0.7\linewidth, angle=-90]{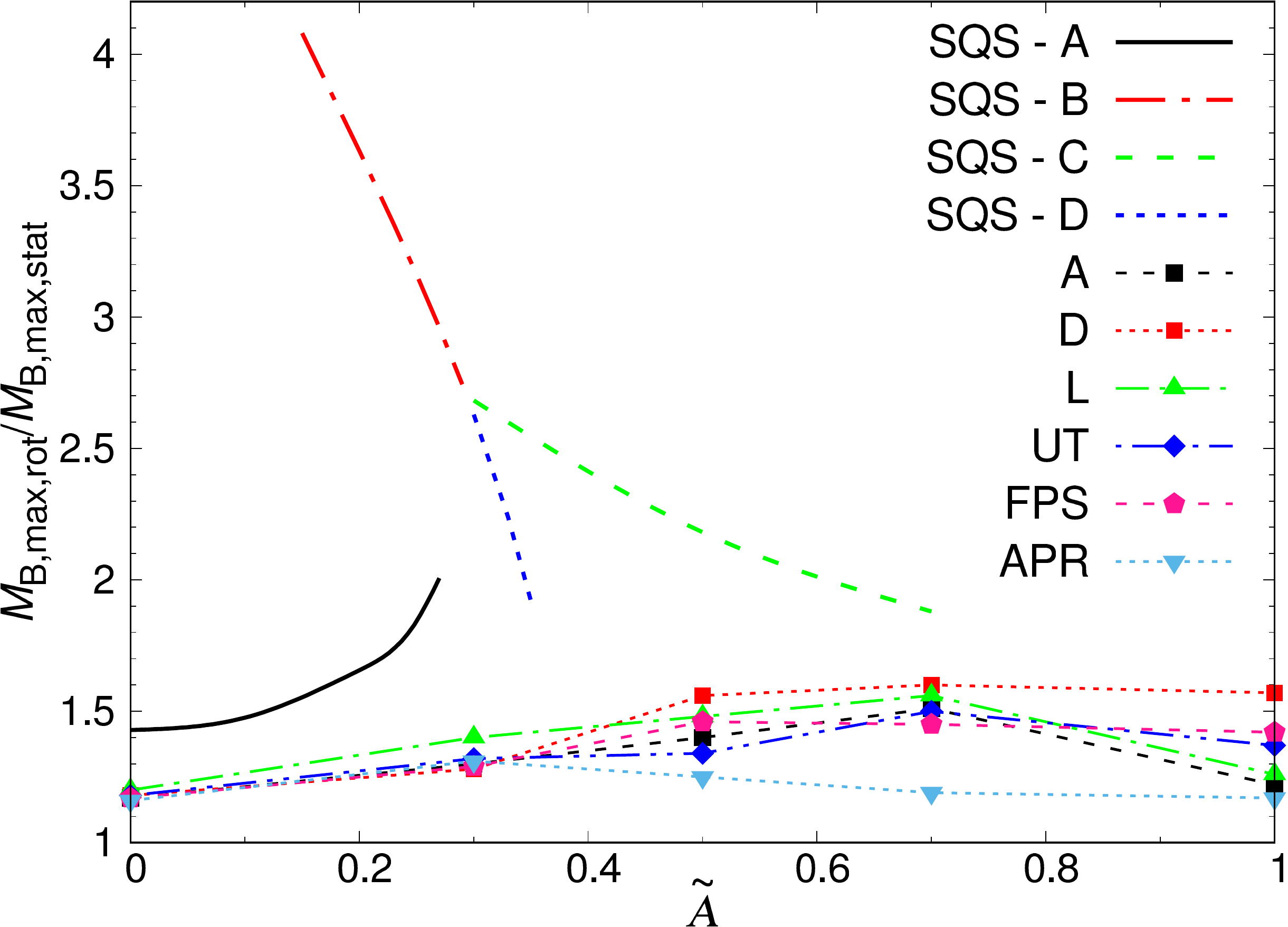}
  \caption{Maximum baryonic mass of differentially rotating strange stars normalized by the maximum mass in the non-rotating case as a function of the degree of differential rotation $\tilde{A}$. The lines without points correspond to our results (Fig.\ref{Fig:M_hmax}), while lines with symbols are for differentially rotating neutron stars described by different realistic equations of states taken from \citet{MBS}.}
  \label{Fig:SQS_NS_Comparison}
\end{figure}

The main results of our study of the maximum mass of differentially rotating strange stars are gathered in Tables~\ref{Tab:Mmax_AC} and~\ref{Tab:Mmax_BD}, as well as on Fig.~\ref{Fig:SQS_NSPoly_Comparison} and~\ref{Fig:SQS_NS_Comparison}. On Fig.~\ref{Fig:SQS_NSPoly_Comparison}, we compare the relative maximum mass increase of SQS with its values for polytropes with three different $\Gamma$ indices [results taken from \citet{GKVAK} - blue lines for $\Gamma = 2.0$ - and \citet{SKGVA} - red and green lines for $\Gamma = 1.8, 2.5$, respectively]. We see that for SQS, the maximum mass depends on both the degree of differential rotation and on the type of solution, as was the case for neutron stars described by polytropic equations of state. However, for strange stars, types B, C and D occur at much smaller degrees of differential rotation $\Atil$ than for polytropes. Additionally,the maximum mass is an increasing function of the degree of differential rotation $\Atil$ when the sequence is of type A (with a highest mass close to the mass-shedding limit), but it becomes a decreasing one once $\Atil$ is sufficiently large for the sequences to be of type C (in which case the maximum mass is reached at the entrance in the toroidal regime, for $\rrat=0$, and is a supremum but not an actual maximum), or of types B and D (both with a maximum mass at the mass-shedding limit). Taking into account simultaneously those two behaviors, the intuitive conclusion is challenged since the largest increase of the maximum mass of differentially rotating strange stars is obtained for a low degree of differential rotation $\Atil$ and not for the highest ones. More quantitatively, for all values of $\Atil$ considered in this study, the largest increase was found for $\Atil=0.15$, giving a configuration with a baryonic mass $M_\mathrm{B} = 10.81 \ \mathrm{M_\odot}$ (an increase of around 308\% with respect to the non-rotating case), a gravitational mass $M = 7.59 \ \mathrm{M_\odot}$, a circumferential radius $R_\mathrm{circ} = 30.00 \ \mathrm{km}$ and a maximum density $\rho_\mathrm{max} = 0.46 \cdot 10^{15} \mathrm{g} \ \mathrm{cm}^{-3}$ (see Table~\ref{Tab:Mmax_BD} for more precise values and additional properties). Such an increase of the maximum mass allowed by differential rotation, for small values of $\Atil$, is much larger than what was found for neutron stars described by polytropic EOSs in previous studies. If we assume that GW170817 merger have led to the quite quick formation of a black hole \citep{Rezzolla2018ApJ}, such a high value is of course questionable. Direct natural conclusions could be that quark matter does not exist, or that differential rotation is dissipated on a very short timescale. However, one should also not forget that our code does nothing more than finding equilibrium configurations: it does not allow us to say anything about their dynamical stability. As a consequence, one could still reasonably think that maybe GW170817 hosted a phase transition from nuclear to quark matter, as proposed in \citet{D2016, 2019PhRvL.122f1101M}, but that it was very fast followed by the collapse of the stellar corpse, even though its mass was smaller than the highest possible one. Only a deeper analysis based on dynamical numerical simulations with realistic microphysics inputs can solve this question. In addition we should keep in mind that the picture on the fate of the post-merger object is still uncertain. For example taking  into account electromagnetic observations of GW170817 the long-lived massive neutron star surrounded by a torus was used \citep{2017SF} to interpret the data or short lived high mass neutron star \citep{MM2017,SJ2019}.

\begin{deluxetable*}{cllllllllll}
\tablecaption{Maximum baryonic mass increase (in \%, compared to the static configuration) for differentially rotating stars described by different EOSs: Strange Quark Stars (SQS) and six realistic nuclear matter EOS (A, D, L, UT, FPS, APR) taken from \citet{MBS}. The corresponding value of $\rrat$ for the configuration with maximum mass is indicated between parenthesis. See the precise references of the articles that describe the nuclear EOS in \citet{MBS}. \label{Tab:SQS_NS_Comparison}} 
\tablehead{
    \colhead{} &
    \colhead{$M_{\mathrm{B,max}}$} & 
    \colhead{$\delta M_{\mathrm{B}}$ ($r_{\mathrm{p}}/r_{\mathrm{e}}$)} & 
    \colhead{$\delta M_{\mathrm{B}}$ ($r_{\mathrm{p}}/r_{\mathrm{e}}$)} & 
    \colhead{$\delta M_{\mathrm{B}}$ ($r_{\mathrm{p}}/r_{\mathrm{e}}$)} & 
    \colhead{$\delta M_{\mathrm{B}}$ ($r_{\mathrm{p}}/r_{\mathrm{e}}$)} & 
    \colhead{$\delta M_{\mathrm{B}}$ ($r_{\mathrm{p}}/r_{\mathrm{e}}$)} & 
    \colhead{$\delta M_{\mathrm{B}}$ ($r_{\mathrm{p}}/r_{\mathrm{e}}$)} & 
    \colhead{$\delta M_{\mathrm{B}}$ ($r_{\mathrm{p}}/r_{\mathrm{e}}$)} \\
    \colhead{} &
    \colhead{$[\mathrm{M_\odot}]$} & 
    \colhead{[\%]} & 
    \colhead{[\%]} & 
    \colhead{[\%]} & 
    \colhead{[\%]} & 
    \colhead{[\%]} & 
    \colhead{[\%]} & 
    \colhead{[\%]}
}
\startdata
    EoS & $\Omega = 0.0$ & $\tilde{A} = 0.0$ & $\tilde{A} = 0.1$ & $\tilde{A} = 0.2$ & $\tilde{A} = 0.3$ & $\tilde{A} = 0.5$ & $\tilde{A} = 0.7$ & $\tilde{A} = 1.0$ \\
    \hline
    SQS 	& 2.65 & 43 (0.468) & 48 (0.452) & 263 (0.10) & 174 (0.01) & 118 (0.01) & 88 (0.01)  & - \\
    A\,\tablenotemark{a}	& 1.92 & 17 (0.565) & -          & -          & 30 (0.46)  & 40 (0.415) & 51 (0.22)  & 22 (0.31) \\
    D\,\tablenotemark{b}	& 1.89 & 18 (0.565) & -          & -          & 28 (0.505) & 56 (0.385) & 60 (0.275) & 57 (2.965) \\
    L\,\tablenotemark{c}	& 3.23 & 20 (0.55)  & -          & -          & 40 (0.425) & 48 (0.32)  & 56 (0.01)  & 26 (0.025) \\
    UT\,\tablenotemark{d}	& 2.17 & 18 (0.565) & -          & -          & 32 (0.46)  & 34 (0.49)  & 50 (0.22)  & 37 (0.01) \\
    FPS\,\tablenotemark{e}& 2.10 & 17 (0.565) & -          & -          & 29 (0.475) & 46 (0.36)  & 45 (0.275) & 42 (0.01) \\
    APR\,\tablenotemark{f}& 2.67 & 16 (0.58)  & -          & -          & 31 (0.445) & 25 (0.37)  & 19 (0.28)  & 17 (0.01) \\
\enddata
    \tablenotetext{a}{A - Reid soft core \citep{P}.} 
    \tablenotetext{b}{D - Model V \citep{BJ}.}
    \tablenotetext{c}{L - Mean field \citep{PS}.}
    \tablenotetext{d}{UT - UV14 + TNI \citep{WFF}.}
    \tablenotetext{e}{FPS - UV14 + TNI \citep{LRP}.}
    \tablenotetext{f}{APR - A18 + $\delta\upsilon$ + UIX* \citep{akmal}.}
\end{deluxetable*}

A first step towards realistic calculations consists in checking whether there is such a difference between quark stars and neutron stars when these latter are also modeled using realistic EOSs. A first comparison between realistic differentially rotating neutron stars and quark stars can be done using our results and those of \citet{MBS} (who studied the influence of various realistic nuclear EOSs on the maximum mass of differentially rotating neutron stars using the same rotation law as us), which is done for the maximum baryonic mass in Table \ref{Tab:SQS_NS_Comparison} and on Figure~\ref{Fig:SQS_NS_Comparison}.

More precisely, in the Table we gathered, for stars with various degree of differential rotation $\Atil$, the relative increase of the maximum baryonic mass with respect to the maximum static mass $(\Omega=0)$. Also, we indicate between parenthesis the corresponding value of $\rrat$. 
As can be seen, we were actually able to compare neutron and quark stars for four values of the degree of differential rotation $\Atil$ (one of which is the case of rigid rotation $\Atil = 0$) as the work of \citet{MBS} did not include $\Atil = 0.1$ nor $0.2$, while we did not study $\Atil = 1$.

Table~\ref{Tab:SQS_NS_Comparison} shows that for each fixed degree of differential rotation the highest increase of the maximum mass, comparing to non-rotating configurations, is reached by strange quark stars, as was already known for rigid rotation. However, in \citet{MBS}, there is a graph (Fig.~2) which is quite analogue to our Figure~\ref{Fig:M_hmax}, and which depicts the highest mass as a function of the maximum density for fixed $\Atil$. Comparing those figures, and taking also into account the results presented for polytropes in \citet{SKGVA}, it seems reasonable to conclude that while for low degrees of differential rotation the code of \citet{MBS} did allow the determination of the actual maximum mass, things are not so clear for high $\Atil$. Indeed the curves they present give the impression that the authors did not get proper sequences of configurations for high degrees of differential rotation and high maximal energy densities ($\epsilon_\mathrm{max}$), the actual maximum mass being probably larger. This assumption is supported by the fact that the critical value of $\Atil$ is around $0.5$ or $0.6$ for polytropes with adiabatic indices similar to the effective indices calculated by \citet{MBS} ($\Gamma_\textrm{eff}\sim 2.5$ or $3$), as shown in \citet{SKGVA}. This implies that the sequences with high $\Atil$ should be of type C or D and the maximum mass is {\it a priori} found for configurations with $\rrat\sim0$ that their code could maybe not calculate precisely. All this would suggest that even if the relative increase of the maximum mass is in all likelihood larger for SQSs, the quantitative difference with its value for realistic models of neutron stars is still an open question.

\section{Conclusions}\label{sec:conc}

We presented the first study of the maximum mass of differentially rotating relativistic stars made from quark matter, using the MIT Bag model. Our results were obtained with the highly accurate spectral code FlatStar, which allows to determine the structure of strongly deformed configurations. By making internal tests and by comparing our code with others in the public domain (RNS and LORENE), we showed that its accuracy is indeed quite high and that it enables a thorough and precise exploration of the solution space for broad ranges of the degree of differential rotation $\Atil$ and of the maximum density $\rhm$.

We built numerous sequences of configurations with fixed $\Atil$ and $\rhm$, and observed that depending on the value of $\Atil$, as was already discovered for homogeneous stars and polytropes in \citet{AGV,GKVAK} and \citet{SKGVA}, there can be several types of sequences with different properties. There is indeed a critical value $\Atc(\rhm)$ such that a sequence starting from a non-rotating configuration finishes at the Keplerian limit if $\Atil<\Atc$, but enters into the toroidal regime if $\Atil>\Atc$. Also, again as was the case for constant density stars and polytropes, we found two threshold values of $\Atil$, noted $\Atil_B$ and $\Atil_D$, such that sequences without a non-rotating limit (said of B or D type) can appear. More precisely, we have the following restrictions of the possible coexistence of the various types:

\begin{itemize}
   \item $0\le\Atil<\Atil_B$ - type A only;
   \item $\Atil_B<\Atil<\Atc$ - type A or B;
   \item $\Atc<\Atil<\Atil_D$ - type C or D;
   \item $\Atil_D<\Atil$ - type C only.
\end{itemize}

We find that the maximum mass of SQS depends on both the degree of differential rotation and on the type of solution. It is an increasing function of $\Atil$ for type A solutions and a decreasing one for types B, C and D. For type A sequences, which end at the mass-shedding limit, the configuration with maximum mass was found close to that limit (but not at it). For type C sequences, it was obtained for the toroidal-like configurations with a vanishing ratio between the polar and equatorial radii. Since such sequences were arbitrarily terminated at that point, the conclusion is that in that case the maximum mass is just a supremum and even higher values could be obtained for non-simply connected configurations. For type B and D, which exists only for small values of $\rrat < 0.25$, the higher mass configurations were found at the mass-shedding limit. Similar behaviors were previously observed for polytropic equations of state \citep{GKVAK, SKGVA}.

Gathering all results, it turned out that the largest increase of the maximum mass with respect to its value in the non-rotating case is reached for type B sequences. This conclusion is again the same as for neutron stars described by a polytropic EOS  \citep{GKVAK, SKGVA}.  We found that the maximum mass of differentially rotating SQS is obtained for a relatively low degree of differential rotation, $\tilde{A}\sim 0.15$, and could even be four times higher than the maximum mass of non rotating strange quark stars described by the same EOS. This increase is much larger than for neutron stars modeled with realistic EOSs as was shown by comparing our results with those of \citet{MBS}. A natural extension of our work would be to use the FlatStar code for realistic nuclear EOSs in order to explore the corresponding solution space in detail and make the comparison with strange matter more precise.\footnote{Recently \citet{EP} confirmed that the universal relation, discovered by \citet{GKVAK}, between maximum mass and the degree of differential rotation is also valid for realistic EOS. Our conclusion that for  small degrees of differential rotation the maximum mass of SQS is much larger than the maximum mass for differentially rotating NS still holds taking into account the results of \citet{EP}.} Also, another important issue, to be dealt with using dynamical simulations, is the stability of massive differentially rotating stars that could be formed during core collapse or in mergers of neutron stars binaries. Indeed, knowing the theoretical life span of such bodies is crucial to put constraints on observations that could enable to demonstrate the existence of quark matter in the Universe and also to better understand the last stages of the dynamics of compact binaries. For instance, even if some analyses of GW170817 suggest a quick collapse into a black hole \citep{Rezzolla2018ApJ, Ruiz2018PRD, Shibata2017PRD}, it cannot be ruled out that it was maybe predated by a transition from nuclear to quark matter \citep{2019PhRvL.122f1101M}.

\acknowledgments
This work is dedicated to the memory of our deeply regretted friend and colleague Marcus Ansorg who sadly passed away before this article could be finished. Nothing could have be done without him. We also would like to thank the anonymous referee for their relevant comments. This work was supported by National Science Center grant UMO-2015/17/N/ST9/01605;  by POMOST/2012-6/11 Program of Foundation for Polish Science co-financed by the European Union within the European Regional Development Fund, by the grant of the National Science Centre UMO-2014/14/M/ST9/00707 and DEC-2013/08/M/ST9/00664, by the COST Actions MP1304, CA16104, CA16214 and by the HECOLS International Associated Laboratory programme. Calculations were performed on the PIRXGW computer cluster at University of Zielona Gora funded by the Foundation for Polish Science within the FOCUS program.

\software{FlatStar for rigid rotation \citep{AKM,AKMb,SA,A,AP},
    FlatStar for differential rotation \citep{AGV,GKVAK},
    RotSeq \citep[][\href{http://www.lorene.obspm.fr}{http://www.lorene.obspm.fr}]{GLorene},
    RNS \citep{SF95}}

\bibliography{bib_file}

\end{document}